\documentclass[twocol]{ametsocV6.1}

\usepackage{booktabs}
\usepackage{physics}
\setcitestyle{notesep={; }}

\usepackage{xspace}
\newcommand{\ie}{\textit{i}.\textit{e}.,\@\xspace}
\newcommand{\eg}{\textit{e}.\textit{g}.,\@\xspace}
\newcommand{\vs}{\textit{vs}.\@\xspace}

\title{The Effects of Spatial Interpolation on a Novel, Dual-Doppler 3D Wind Retrieval Technique}

\authors{Jordan P. Brook \aff{a}\correspondingauthor{Jordan Brook, j.brook@uq.edu.au}
Alain Protat,\aff{b} 
Corey K. Potvin,\aff{c,d} 
Joshua S. Soderholm,\aff{b} 
Hamish McGowan,\aff{a} 
}

\affiliation{\aff{a}{Atmospheric Observations Research Group, University of Queensland, St Lucia, QLD, Australia}\\
\aff{b}{Science and Innovation Group, Australian Bureau of Meteorology, Docklands, VIC, Australia}\\
\aff{c}{NOAA/OAR/National Severe Storms Laboratory, Norman, Oklahoma}\\
\aff{d}{School of Meteorology, University of Oklahoma, Norman, Oklahoma}}

\abstract{Three-dimensional wind retrievals from ground-based Doppler radars have played an important role in meteorological research and nowcasting over the past four decades. However, in recent years, the proliferation of open-source software and increased demands from applications such as convective parameterizations in numerical weather prediction models has led to a renewed interest in these analyses. In this study, we analyze how a major, yet often-overlooked, error source effects the quality of retrieved 3D wind fields. Namely, we investigate the effects of spatial interpolation, and show how the common practice of pre-gridding radial velocity data can degrade the accuracy of the results. Alternatively, we show that assimilating radar data directly at their observation locations improves the retrieval of important dynamic features such as the rear flank downdraft and mesocyclone within supercells, while also reducing errors in vertical vorticity,  horizontal divergence, and all three velocity components.}

\begin{document}


\maketitle

\addtolength{\topmargin}{-44.23164pt}


%
%
%
\statement{We can attempt to estimate the wind speed and direction within a weather system when two weather radars measure it simultaneously. However, radars do not scan the whole atmosphere at once - instead, they measure along many cross sections, each at different heights. We show that a method commonly used to stitch the observations together degrades the accuracy of the winds. Additionally, we describe a way to feed the data directly into the analysis without stitching it together first, and show that this improves the wind retrievals considerably. We hope these improvements will help researchers better understand how various weather systems work, and help forecasters warn for dangerous weather such as tornadoes. }

%


\section{Introduction}\label{intro}

Three-dimensional wind retrievals from Doppler radar measurements have become an indispensable tool for studying the kinematic properties of various weather systems, such as tropical cyclones, convective storms and tornadoes \citep[\eg][]{Kosiba14, Betten18, Markowski18}. These analyses are also set to play an important role in observationally verifying and improving numerical weather prediction model simulations and parameterizations of convection \citep[\eg][]{Kumar15, Nicol15, Labbouz18}. Additionally, 3D wind retrievals herald exciting opportunities for severe weather forecasting, \eg for damaging surface winds and updraft intensity identification, or other research applications such as real-time hail trajectory nowcasting and growth modelling \citep{Kumjian21, Brook21}. With these applications in mind, this study aims to investigate how the commonly overlooked effects of spatial interpolation impact current wind retrieval algorithms. Our study also particularly focuses on 3D wind retrievals for operational radar networks \citep[\eg][]{Bousquet07, Dolan07, Park09b} where the undesirable effects imposed by spatial interpolation are most severe.\footnote{Spatial interpolation errors are most severe in operational networks due to the large horizontal distances ($>50$ km) common between radars, and large elevation gaps (5$^\circ$--6$^\circ$) between constant elevation scans in operational scanning strategies.} We also expect our findings to be useful in retrievals using airborne, or rapid-scan mobile radar data. 

It is now the literary consensus that variational 3D wind retrievals outperform those involving numerical integration of the mass continuity equation \citep{Gao99, Potvin12, North17}. These improvements are attributed to well-known methodological shortcomings of the latter, involving ill-posed boundary conditions and vertically compounding errors during numerical integration \citep{Ray80}. Despite the improvements offered by variational analyses, the retrieval of accurate 3D winds from typical dual-Doppler observations remains a challenging problem. These challenges are greatest for the vertical velocity component, due to the focus on low elevations in typical operational scanning strategies. Under these conditions, the vertical component of the 3D velocity vector is poorly observed, meaning vertical velocity retrievals must rely heavily on the mass continuity constraint. Furthermore, the task is complicated by a number of secondary factors, including spatial interpolation errors, data gaps near the ground and near the storm top, measurement non-simultaneity, discretization errors and erroneous Doppler velocity data \citep[\eg due to incorrect dealiasing, dual-PRF artefacts, side lobe contamination or ground clutter;][]{Gao99, Potvin12b}. All of these factors contribute to the considerable errors noted in variational 3D wind retrievals, especially in the vertical velocities \citep{Potvin12d, Oue19}.

\citet{Gao99} summarized that the two most pronounced difficulties for vertical velocity retrievals are: 1) data gaps at the vertical domain boundaries, and 2) the spatial interpolation of radar information. In this study, we address the impacts of the latter, and aim to asses how errors introduced by spatial interpolation propagate into variational 3D wind retrievals. However, we first pause to note the significant progress made on the first of these issues. Perhaps the most notable contribution was made by \citet{Shapiro09}, who showed that including the vertical vorticity equation as a variational constraint can significantly improve retrievals with data voids near the surface in observing system simulation experiments (OSSEs) of simple analytic fields. \citet{Potvin12b} consolidated this work by showing this constraint also aids vertical velocity retrievals in low-level data denial OSSEs of simulated thunderstorms, and this finding has been reproduced in more recent studies \citep{Dahl19, Gebauer22}. In practice, the provision of a vertical vorticity constraint requires data from multiple consecutive radar scans to estimate its temporal evolution \citep[\eg][]{Protat00}. This can be achieved to varying degrees of accuracy by estimating a constant horizontal advection vector \citep{Shapiro09}, or by using a combination of provisional retrievals at successive time steps and spatially-varying advection correction techniques \citep{Potvin12b, Dahl19, Gebauer22}. These studies, along with others such as \citet{Potvin12d} and \citet{Oue19}, have shown that large errors can arise when the scanning time of a radar volume approaches that of operational weather radars ($\sim$5 min.) due to erroneous temporal discretizations and temporal evolution of wind fields. In practice, the interpolation practices discussed here should be used in conjunction with these previous advances.

Spatial interpolation procedures for radial velocity data in 3D wind retrievals may be characterized into two types: 1) pre-gridded methods, and 2) direct radar assimilation methods. The former requires that radar data is interpolated to a common, Cartesian analysis grid prior to the analysis. This simplifies the observational constraint, but introduces significant errors in radial velocities before the variational retrieval has begun \citep{Gao99, North17, Oue19}. Gridding methods used in wind retrievals vary from simple nearest neighbour/linear interpolation methods such as that in \citet{Mohr79} \citep[used in][]{Sun98, Protat99, Collis10}, weighted averages such as \citet{Cressman59} or \citet{Barnes64} \citep[used in][]{North17, Oue19, Gebauer22}, or a combination of the two in \citet{Dahl19}. The reader is referred to \citet[][hereafter B22]{Brook22} for a comprehensive discussion on the types of errors introduced by the various radar gridding techniques. The propagation of gridding errors into 3D wind retrievals has been studied by \citet{Collis10} for linear interpolation and \citet{Majcen08} for multi-pass Barnes interpolation. However, it is difficult to draw reliable conclusions regarding the relative performance of gridding methods between these studies, due to the differences in study setups (analytical OSSEs \vs NWP model OSSEs), weather types (simple horizontal flow \vs supercell) and varying retrieval methodologies/software implementations. To the authors' knowledge, no effort has yet been made to provide an inter-comparison of multiple gridding methods and their effects on 3D wind retrievals.

The second category of spatial interpolation procedures for 3D wind retrievals is the direct radar assimilation approach. In these methods, radar data is ingested into the observational constraint in its native spherical coordinates, forgoing the pre-gridding of radial velocities. A comparison of the radial velocities at their measurement locations is facilitated in the observation constraint by either a trilinear \citep[\eg][]{Testud83, Gao99, Gao04} or a Cressman/Barnes weighted average \citep[][]{Shapiro09, Potvin12b} operator. \citet{Gao99} notes this ``reverse'' interpolation from the regularly-spaced analysis grid to the irregularly-spaced radar locations is naturally well-defined, whereas interpolation from the radar locations to the analysis grid (as in standard radar gridding) is often ill-posed due to the large data voids between tilts in operational radar data. Encouragingly, \citet{Potvin12d} found that the analysis was largely insensitive to the Cressman radius of influence used in their forward operator, an indication of the well-posed nature of this type of assimilation procedure. The radar assimilation method also benefits from performing the interpolation simultaneously with the other variational constraints, such that the mass continuity and smoothness penalties also assist in the interpolation procedure. Additionally, the exact radial contribution of each Doppler measurement is also preserved in the radar assimilation method, avoiding the spatial distortion involved with averaging many individual measurements as in the pre-gridded radial velocity data. These methodological advantages lead us to hypothesize that the radar assimilation method should result in smaller spatial interpolation errors relative to the pre-gridded retrieval methods \citep{Gao99, Rihan05}. We aim to verify this hypothesis by directly comparing pre-gridded 3D wind retrievals to those using direct radar assimilation. 

We focus our analysis on supercells in this study for two reasons: 1) whilst they represent a small fraction of global convection, they are disproportionately responsible for severe weather hazards \citep[\eg][]{Duda10, Smith12}, and 2) they present the greatest challenge for current wind retrieval methodologies due to the presence of complex, meso-scale circulations and strong vertical motions \citep{Potvin12b, Shapiro09}. The remainder of the manuscript is organized as follows; firstly, we introduce the OSSE and wind retrieval methodologies used to test our hypothesis in Section \ref{methods}. Secondly, we present qualitative and quantitative results from our ensemble of supercell OSSE experiments in Sections \ref{results}\ref{ss:case} and \ref{results}\ref{ss:ensemble}, before assessing the applicability of these findings on real supercell data in Section \ref{results}\ref{case}. Finally, we summarize our findings and discuss future research directions in Section \ref{summary}.


\section{Methodology}\label{methods}

\subsection{OSSE Setup}\label{OSSE}

\begin{figure}[t]
\centering
\includegraphics[width=0.5\textwidth]{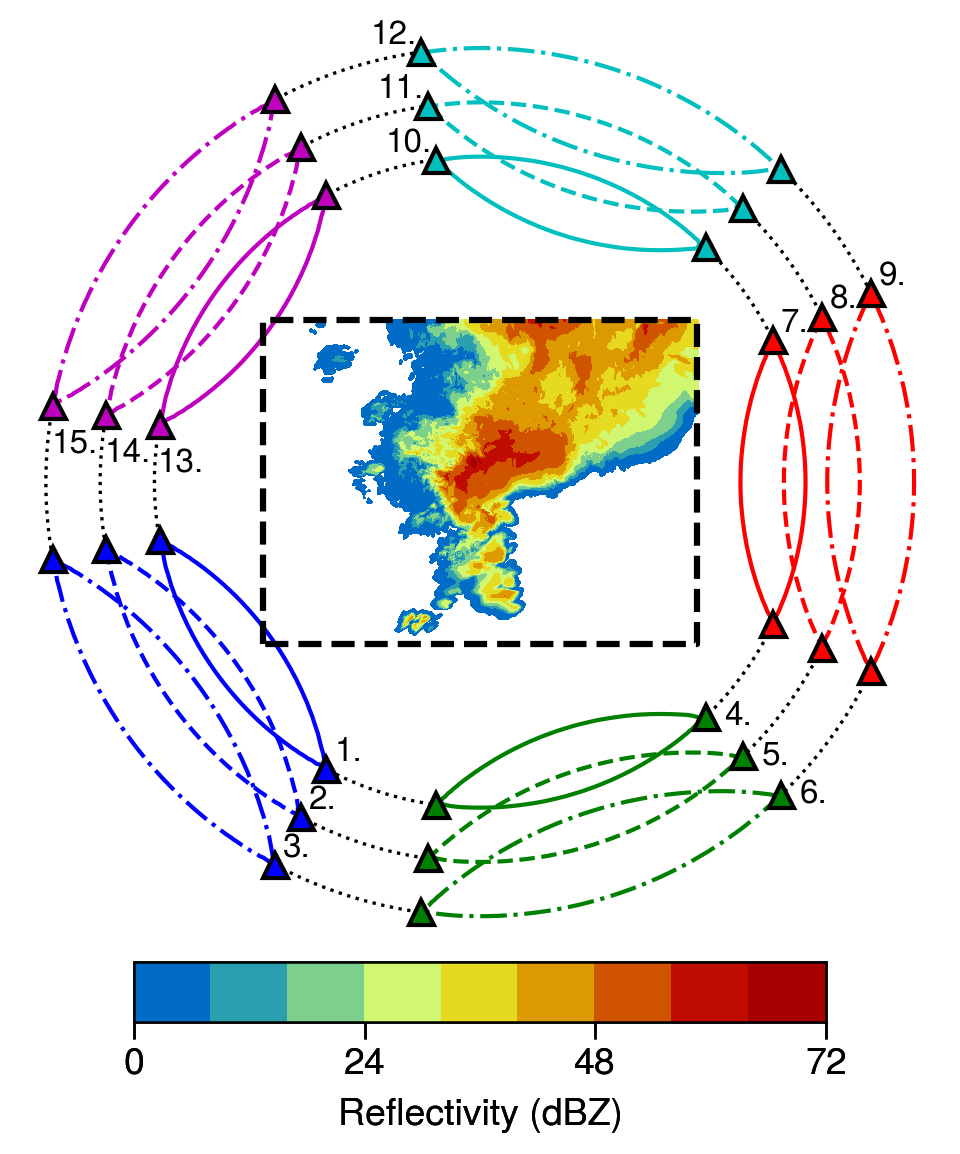}\\
 \caption{\label{setup} The OSSE ensemble setup for this study. Dual-Doppler data are simulated for fifteen pairs of locations around the model grid (pictured with reflectivity at the 20 m model level). Each simulated radar position is represented by a triangle, with colored lines indicating the collinear beam region between each radar pair (cosine of cross beam angle $>$ 0.9). The ensemble is made up of five different viewing angles (216$^\circ$, 288$^\circ$, 0$^\circ$, 74$^\circ$ and 144$^\circ$, clockwise from the positive $x$ axis, shown by varying colors) and three different ranges (60 km, 70 km and 80 km from the center of the model grid, shown by varying line styles).}
\end{figure}

In this study, we examine the effects of spatial interpolation on 3D wind retrievals using observing system simulation experiments \citep[OSSEs, \eg][]{Potvin12, Potvin12b, Dahl19}. Under the OSSE framework, observations are simulated using a known source, meaning the accuracy of the resulting experiments may be quantified exactly (in contrast to experiments with real data, where the ``truth'' is unknown). The source of our OSSE data is an Advanced Regional Prediction System \citep[ARPS,][]{Xue00} simulation of the 8 May 2003 Oklahoma City tornado, first described in \citet{Xue14}. The simulation contains a mature, nearly steady-state tornadic supercell shown in Fig.\ \ref{setup}. The reader is referred to \citet{Xue14} and \citet{Schenkman14} for a detailed description of the supercell. The ARPS model contains a terrain-following grid with 53 vertical levels, increasing from $\sim$20 m spacing at ground level to $\sim$750 m spacing at the top of the domain, and a uniform 50 m spacing in the horizontal dimensions. The simulation begins at 2200 UTC and extends $20\times60\times80$ km in the $z$, $y$ and $x$ dimensions, respectively, having been successively downscaled through a series of three nested domains with 9 km, 1 km and 100 m horizontal grid spacings. Mesonet, rawinsonde and Weather Surveillance Radar-1988 Doppler (WSR-88D) data are all assimilated into the analysis, and the reader is referred to \citet{Xue01} and \citet{Xue03} for additional details on the model physics. 

Modelled wind fields from the 20th minute of the 50 m resolution ARPS simulation (2220 UTC) serve as the ``true'' atmospheric state\footnote{This analysis time is identical to that in \citet{Dahl19}, meaning retrieved winds may be qualitatively compared to their findings despite the smaller analysis domain used in their study.}, upon which radar data is emulated at various positions around the domain. More specifically, our OSSE experiments contain an ensemble of fifteen different dual-Doppler observation locations (Fig.\ \ref{setup}). The azimuthal spacing between radar pairs is set to 51.7$^\circ$, so that the model grid lies at the center of the respective dual-Doppler lobe in each experiment (defined where the cosine of the cross-beam angle $<$ 0.9, shown as black dotted lines in Fig.\ \ref{setup}), eliminating errors due to poor cross-beam angles. We implement the experimental ensemble to reflect the various sampling geometries possible in operational domains, thereby ensuring the generality and repeatability of the findings. We choose to extend our analysis grid from the lowest grid point at 500 m above sea level (ASL, note the terrain varies gradually between $\sim$270--390 m ASL across the domain), up to the tropopause height ($\sim$15 km altitude) and over the entire ARPS domain (60 km meridionally and 80 km zonally). The chosen domain is significantly larger than previous rapid-scan OSSE experiments \citep[\eg the $5\times20\times20$ km analysis grid in][]{Dahl19}. Radars are positioned at an altitude of 350 m, either 60, 70 or 80 km from the center of the domain (illustrated by the varying line styles in Fig.\ \ref{setup}), resulting in baseline distances for dual-Doppler pairs of $\sim$50, $\sim$60 and $\sim$70 km, respectively. Furthermore, the farthest range for the most distant radars in our experiment is roughly 130 km. These settings reflect our focus on retrieval methodologies applicable to operational radar networks (as opposed to small-scale, rapid-scan analyses with mobile radars), where dual-Doppler baselines are commonly large\footnote{For example, the largest dual-Doppler baseline distance tested in this study corresponds to the distance between the two operational radars (70 km, Terrey Hills and Wollongong) in Sydney, Australia's most populous city.}, and analysts wish to retrieve winds over large domains. 

Radar data simulation is implemented identically to previous studies \cite[\eg][B22]{Dahl19}, and is described here for completeness. The radar beam propagates assuming standard atmospheric refraction, and accounts for the curvature of the earth using the 4/3 effective earth radius model \citep{Doviak93}. Model points within each radar voxel are weighted according to their proximity to the center of the beam, effectively functioning as individual scatterers. Radar observations are calculated as a weighted average of each scatterer as follows,

\begin{equation}\label{weights}
    \varphi = \frac{\sum_{i=1}^n R_iB_i\varphi_i}{\sum_{i=1}^n R_iB_i},
\end{equation}

\noindent
where $\varphi$ is the radar field (either radial velocity or reflectivity in our case), $R_i$ and $B_i$ are radial and radar beam weights for the $i^{\text{th}}$ of $n$ scatterers within each radar voxel. The resolution of the underlying model data ensures each radar measurement is informed by at least one ``scatterer''. Radial weights are implemented as,

\begin{equation}
\label{r_weight}
	R_i =
    \begin{cases}
        1, \hfill \abs{\Delta r_i}<0.3d \vspace{0.1cm}\\
        \max\left(\frac{0.5d-\abs{\Delta r_i}}{0.2d}, 0 \right), \qquad \abs{\Delta r_i}\geq 0.3d
    \end{cases},
\end{equation}

\noindent 
where $\Delta r$ is the scatterer's radial offset from the center of the radar voxel and $d=250$ m is the radial data spacing. As in \citet{Potvin09}, the following beam weighting function simulates the angular sensitivity of radars within the United States operational radar network (WSR-88D), 

\begin{equation}\label{b_weight}
    B_i = \exp{-8\ln(2)\left[\left(\frac{\Delta \theta_i}{\theta_B}\right)^2 + \left(\frac{\Delta\phi_i}{\phi_B}\right)^2 \right]},
\end{equation}

\noindent 
where $\Delta \phi_i$  and $\Delta \theta_i$ are the azimuthal and elevation offsets and $\phi_B = \theta_B= 1^\circ$ are the half-power beamwidths (chosen to mimic standard operational S-band radars). We also implement the operational scanning strategy used in the Australian radar network, which contains 14 Plan Position Indicator (PPI) sweeps at the following elevations: $\theta =$ (0.5, 0.8, 1.4, 2.4, 3.5, 4.7, 6.0,  7.8, 10.0, 13.0, 17.0, 23.0, 32.0, 45.0)$^\circ$. The realistic range, scan strategy and dual-Doppler baselines used in these experiments ensure the applicability of these results to real, operational radar data. To this end, we follow B22 by contaminating all Doppler simulations with Gaussian observational noise (standard deviation of 1.0 m s$^{-1}$) and randomly masking 5\% of all measurements to mimic measurement errors. 

Radar reflectivity is calculated based on the modeled hydrometeor mixing ratios according to \citet{Tong05}, and all Doppler measurements coincident with $<$5 dBZ reflectivity are masked to mimic data gaps in non-precipitating regions 
\citep[\eg][]{Potvin12d}. Finally, we assume all radar information is collected instantaneously to eliminate temporal discretization effects due to flow evolution and translation, and do not include any consideration on how hydrometeor terminal velocities affect the measured Doppler velocities (\ie we calculate radial wind velocities directly from the modeled $u$, $v$ and $w$ wind components assuming particle velocity is nil). While both error sources must be accounted for in retrievals from observations [\eg using a vertical vorticity constraint and advection correction procedure \citep{Shapiro09, Potvin12b, Shapiro10a}, along with a reflectivity-based terminal velocity correction \citep{Atlas73}], we control for these effects in our experiments to isolate the errors due to spatial interpolation alone. Wind retrieval outputs from all methods are masked using the 5 dBZ threshold from the true model reflectivity to permit the direct comparison of error statistics. These statistics include root mean squared errors of various quantities, including the wind components themselves, the total wind magnitude, and derived products such as vertical vorticity $\zeta \equiv \partial v/\partial x - \partial u /\partial y$ and horizontal divergence $\delta \equiv \partial u/\partial x + \partial v /\partial y$. Consistent with previous studies \citep[\eg][]{Potvin12d}, centered finite derivatives with Neumann boundary conditions were used for these derived fields. Fractions Skill Scores (FSS) are also calculated on column-maximum, or horizontal slices of vertical velocity fields, to quantify how well each method retrieves updraft/downdraft regions \citep{Roberts08}. FSS scores are well-suited for assessing the ability to retrieve large vertical motions when slight spatial offsets are acceptable, owing to its original use as a neighborhood verification method for NWP rainfall accumulations.\footnote{We average all FSS calculations over five spatial scales in our study (namely 1, 2, 3, 4 and 5 km), and found that our results were insensitive to shortening, lengthening or adding additional length scales (not shown).} Such cases may involve operational nowcasting or informing convection parameterizations, where identifying the size and strength of updrafts/downdrafts may be the priority. Finally, backward trajectory calculations for retrieved wind fields in Section \ref{results}\ref{ss:case} were calculated similarly to \citet{Potvin12d} by using a second-order Runga-Kutta scheme with an adaptive time step, along with trilinear interpolation for the surrounding velocity values.

\subsection{Retrieval Methodology}\label{retrieval_method}

The variational 3D wind retrieval method proceeds with a cost function similar to those in previous literature, albeit with two important distinctions: an altered observational constraint ($J_o$) to facilitate the various interpolation methodologies, and a novel denoising constraint ($J_d$). The overall cost function is given below,

\begin{equation} \label{ret_cost}
    \displaystyle{\min_{\pmb{v}}} \left\{J_o(\pmb{v})+ J_m(\pmb{v}) + J_s(\pmb{v}) + J_{d}(\pmb{v})\right\},
\end{equation}

\noindent
where $\pmb{v} = (u, v, w)$ is the 3D velocity vector, $J_m$ is the mass continuity constraint and $J_s$ is the smoothing constraint, each of which are outlined below. Firstly, in this study, the observational constraint in Eq. \ref{ret_cost} varies depending on the structure of the input radial velocity data ($V_r$). We draw a distinction between methods that require pre-gridded radial velocity data and the proposed radar assimilation (RA) method, which ingests radial velocity data in its native, spherical coordinate system. We outline these two distinct observational constraints below in Sections \ref{methods}\ref{retrieval_method}\ref{pre_grid} and \ref{methods}\ref{retrieval_method}\ref{DA}, before introducing the remaining three variational constraints that are common to all experiments in Section \ref{methods}\ref{retrieval_method}\ref{other_constraints}.

\subsubsection{Pre-Gridded Observational Constraint}\label{pre_grid}

When radar data is pre-gridded onto the analysis grid (the gridding techniques used here are outlined in Section \ref{methods}\ref{gridding}), no spatial interpolation operator is required in the optimization problem. Instead, the three Cartesian wind fields are iteratively compared to radial velocity observations on the $M = n_z\times n_y \times n_x$ sized analysis grid, through a gridded projection operator $\mathcal{P}_g: \mathbb{R}^{3M} \to \mathbb{R}^M$. The projection from Cartesian to radial velocities is achieved through the standard geometric relationship $V_r = \pmb{v} \cdot \bm{p}$, where $\bm{p}$ is the normalized displacement vector between the radar measurement position and the location of the radar. Under these conditions, the observational constraint takes the following form,

\begin{equation}\label{radar_obs}
    J_o =\left|\left|V_{r,g}- \mathcal{P}_g \pmb{v}\right|\right|_2^2 
\end{equation}

\noindent
where $V_{r,g}$ denotes the pre-gridded radial velocity measurements from both radars on a Cartesian analysis grid, and the common $\ell_2$ (\ie Euclidean) norm  notation: $||x||^2_2 = \sum_{i=1}^Nx_i^2$ is also used throughout the text. 

\subsubsection{Radar Assimilation Observational Constraint}\label{DA}

The direct assimilation method ingests radial velocity observations ($V_r$) directly from the locations they were observed by the radar. The observational constraint in this case is as follows,

\begin{equation}\label{cart_obs}
    J_o =\left|\left|V_r - \mathcal{P}\mathcal{C} \pmb{v}\right|\right|_2^2 ,
\end{equation}

\noindent 
where $\mathcal{P}$ and $\mathcal{C}$ are radial velocity projection and Cressman interpolation operators, respectively. During forward passes, the Cressman interpolation operator is used to interpolate the three Cartesian velocity fields from the analysis grid, to the exact radar observation locations $\mathcal{C}: \mathbb{R}^{3M} \to \mathbb{R}^{3N}$, where $N$ is the cumulative number of observations from both radars. The reader is referred to the Appendix for supplementary information on $\mathcal{C}$. Finally, the radial projection operator (as described in Section \ref{methods}\ref{retrieval_method}\ref{pre_grid}) projects the three Cartesian velocity components at each radar data location into radial velocities $\mathcal{P}: \mathbb{R}^{3N} \to \mathbb{R}^N$. As in \citet{Gao04}, the projection and interpolation operators are combined to form a single operator in our software implementation.

\subsubsection{Remaining Constraints}\label{other_constraints}

The second constraint in Eq. \ref{ret_cost} is the anelastic form of the mass continuity constraint, commonly chosen for its suitability in deep convection \citep{Batchelor53}. This constraint is formalized identically to those in previous studies, using an idealized exponential density profile $\rho = \rho_0\exp(-z/H)$, where $\rho_0$ is the reference density\footnote{Note there is no need to set a reference density value ($\rho_0$) as it is cancelled out in the derivation of Eq. \ref{J_m}.} and $H=10$ km is the scale height of the atmosphere \citep[\eg][]{Shapiro09, Potvin12b}. We express it here in its simplest form,

\begin{equation}\label{J_m}
    J_m = \lambda_m \left|\left|u_x + v_y+ w_z - \frac{w}{H}\right|\right|_2^2,
\end{equation}

\noindent
where $\lambda_m$ is a user-defined weighting constant used to control the relative importance of this constraint, and partial derivatives in each Cartesian dimension are denoted by subscripts (\eg $u_{x} = \pdv{u}{x}$).

B22 found that unmixed, second-order derivatives were a compelling regularization constraint for assimilating radar data into a Cartesian grid. The reader is referred to their Appendix B for a discussion on the benefits and drawbacks of other common smoothing constraints. We follow B22, along with other recent wind retrieval studies \citep{Potvin12b, Dahl19, Gebauer22}, in implementing second-order derivatives for this purpose. Theoretically, this formulation leads to a visually pleasing minimum curvature solution \citep{Briggs74}, while also spreading radar information into data voids due to the 3-point, centered finite difference stencil used for the numerical derivatives. The implementation used here is given below,

\begin{equation} \label{js}
    J_s = \sum_{\varphi \in \pmb{v}}\lambda_{\mathrm{v}}||\varphi_{zz}||_2^2  + \lambda_{\mathrm{h}} \left(||\varphi_{yy}||_2^2 + ||\varphi_{xx}||_2^2\right),
\end{equation}

\noindent 
where $\lambda_{\mathrm{v}}$ and $\lambda_{\mathrm{h}}$ are weighting constants for the vertical and horizontal dimensions, respectively, and double subscripts denote second partial derivatives (\eg $\varphi_{xx} = \pdv[2]{\varphi}{x}$).

Injudicious application of heavy second-order smoothing using Eq. \ref{js} will eliminate observational radar noise at the expense of ``over-smoothing'' velocity fields, whereby meaningful information is filtered from the analysis \citep{Testud83, Potvin12d}. B22 found that the inclusion of an anisotropic total variation denoising constraint, along with a conventional second-order smoothing constraint, is able to efficiently spread information in to data voids, eliminate observational noise and preserve the sharp ``edge'' features that are common in radar observations of deep convection. In this study, we implement this constraint in an effort to better resolve the highly non-uniform, turbulent velocity information that is commonly underestimated in 3D wind retrieval analyses of strong thunderstorms \citep[\eg][]{Potvin12d, Oue19, Evaristo21}. The denoising constraint is implemented as follows,

\begin{equation} \label{jd}
    J_d = \lambda_d \sum_{\varphi \in \pmb{v}} ||\varphi_z||_1 + ||\varphi_y||_1 + ||\varphi_x||_1
\end{equation}

\noindent
where $\lambda_d$ is a tunable weighting parameter and the $\ell_1$ norm notation used here is equivalent to $||x||_1 = \sum_{i=1}^N|x_i|$. Optimization in experiments containing the $\ell_1$ denoising constraint is complicated due to the non-smooth, non-differentiable term in the cost function (arising from the first-order discontinuous absolute value operation in Eq. \ref{jd}). As in B22, the split Bregman optimization method for mixed $\ell_1$--$\ell_2$ norms is used for optimizing the cost function in our experiments \citep{Goldstein09,Ravasi20}. 

The provision of suitable tuning parameters ($\lambda_m$, $\lambda_v$, $\lambda_h$ and $\lambda_d$) is a challenging task in variational wind retrievals \citep{Shapiro09}. Theoretically, parameters should reflect the intended strength of each constraint; however, such judgements are difficult to make \textit{a priori} due to factors such as observational errors and data acquisition gaps. Recent OSSE studies acquire suitable tuning parameters through experimentation, aided by the introduction of non-dimensional tuning parameters that narrow the tuning parameter space \citep{Shapiro09, Potvin12b, Dahl19}. We extend this experimental approach by programmatically retrieving optimal tuning parameters for each experiment using Bayesian optimization \citep[\eg][]{Barth22}. Optimal tuning parameters are obtained by minimizing the RMS error for each OSSE retrieval, ensuring a fair comparison between each experiment. As in the aforementioned OSSE studies, these optimized parameters may then be employed in retrievals using observations. This approach is important when judging the potential performance for different variational formalisms, as optimal\footnote{We attempt to find the optimal parameters for each experiment (by experimentally limiting the parameter space in subsequent runs and running each optimization for 100 iterations), however, the reported parameters may not be truly ``optimal'' due to local, rather than global, convergence.} parameters likely vary significantly for each experimental setup (\eg pre-gridded data may benefit from a reduced smoothing constraint as some smoothing has already taken place during the gridding process). We use the Bayesian optimization routines within \textit{scikit-optmize} for these purposes \citep{Head21}.

\subsection{Radar Gridding} \label{gridding}

Most recent 3D wind retrieval studies \citep[\eg][]{North17, Dahl19, Gebauer22}, along with open-source wind retrieval software \citep[\textit{PyDDA},][]{Jackson19}, require that dual-Doppler wind data must be pre-gridded prior to analysis. In this study, we aim to investigate how this pre-processing step may degrade the resulting wind retrievals relative to the direct assimilation approach described above, and provide guidance on how common gridding techniques perform relative to each other. We also include ``control'' dual-Doppler observations, which are upscaled directly from the original, high-resolution (50 m horizontal) model grid using \citet{Cressman59} weighted average interpolation with a radius of influence equal to the isotropic, analysis grid spacing (500 m). These control experiments will illustrate the baseline performance of our wind retrieval code for ``perfect'' observations, before they are degraded by the spatial interpolation errors in the ``radar sampled'' experiments.\footnote{Note that retrievals with ``perfect'' control observations do not result in wind fields with zero errors due to the effects of discretization errors, ill-posed boundary conditions and model assumptions (such as the anelastic mass continuity approximation) in the retrieval methodology.} Furthermore, we test two common radar gridding techniques involving Cressman weighted averages in two and three dimensions, both of which are detailed below.

\subsubsection{3D Cressman}\label{ss:cressman}

Perhaps the simplest and most commonly used radar gridding methods are the weighted average methods first proposed by \citet{Cressman59} and \citet{Barnes64}. These methods are commonly used prior to variational wind retrieval algorithms \citep[\eg][]{North17, Oue19, Gebauer22}, and operate by weighting all radar observations within a radius of influence ($R$) from each grid point in three dimensions. We test Cressman, rather than Barnes, weightings in this study due to their more prominent use in radar literature \citep{Trapp00}.\footnote{B22 found that this distinction is largely academic, as the Barnes and Cressman methods produce practically indistinguishable results for the purposes of domain-wide radar gridding.} The Cressman weighting function ($W$) is outlined below,

\begin{equation}\label{cressman}
    W(r) = \exp(\frac{R^2-r^2}{R^2+r^2})
\end{equation}

\noindent
where $r$ is the distance between an observation and the corresponding grid point. Despite being initially proposed as an iterative technique, radar analysts commonly implement Eq. \ref{cressman} in a single pass \citep[with some exceptions, \eg][]{Majcen08}. We follow this literary precedent by applying the 3D Cressman method in a single pass, using an open-source implementation \citep[PyART,][]{Helmus16}. As in B22, we set $R$ equal to the largest data spacing in the analysis domain ($R = d_{\text{max}} \approx3 \mathrm{km}$) and refer to the discussion in their Section 2b regarding the provision of this parameter.

\subsubsection{2D Cressman} \label{ss:2dc}

The second gridding technique implemented here was first described by \citet{Dahl19} for the purposes of dual-Doppler wind retrievals, and is subsequently referred to as the 2D Cressman method. Here, a single-pass weighted average with Cressman weights (Eq. \ref{cressman}) is used to map radar data onto a horizontally-regular, Cartesian grid along each conical, two-dimensional, PPI surface. These surfaces are then merged onto a 3D Cartesian grid by linearly interpolating along each vertical column according to,

\begin{equation}\label{linear}
    V_r = V_{r,1} + \frac{z-z_1}{z_2-z_1}\left(V_{r,2}-V_{r,1}\right)
\end{equation}

\noindent
where subscripts 1 and 2 refer to data immediately below, and above (respectively) the grid point at height $z$ in each column. 

Using Cressman interpolation solely along PPI surfaces permits the use of a considerably smaller $R$ value ($\sim$1.7 km), set by the maximum azimuthal data spacing instead of the larger maximum elevation spacing (elevation gaps between PPI's commonly exceed 5$^\circ$ in operational scanning strategies). A decreased $R$ value leads to the retention of small-scale details (on the order of $d_{\text{max}}$), which are commonly filtered or ``over-smoothed'' in the standard 3D Cressman gridding approach \citep[][B22]{Trapp00, Zhang05}. Resolution improvements gained through smaller $R$ values are naturally offset by inaccuracies introduced by the vertical linear interpolation, such as first-order discontinuities in sparsely sampled regions and spectral distortion at poorly resolved wavelengths \citep[][B22]{Askelson00, Trapp00}. 3D wind retrievals are particularly sensitive to this type of spectral broadening due to their reliance on finite differences for numerical derivatives \citep{Testud83}, meaning it is not immediately clear whether the added resolution achieved by the 2D Cressman method is advantageous for our purposes. In this study, we aim to provide some experimental guidance regarding this question.


\section{Results} \label{results}

\subsection{Ensemble Member Case Study}\label{ss:case}

\begin{figure*}[t]
\centering
\includegraphics[width=0.93\textwidth]{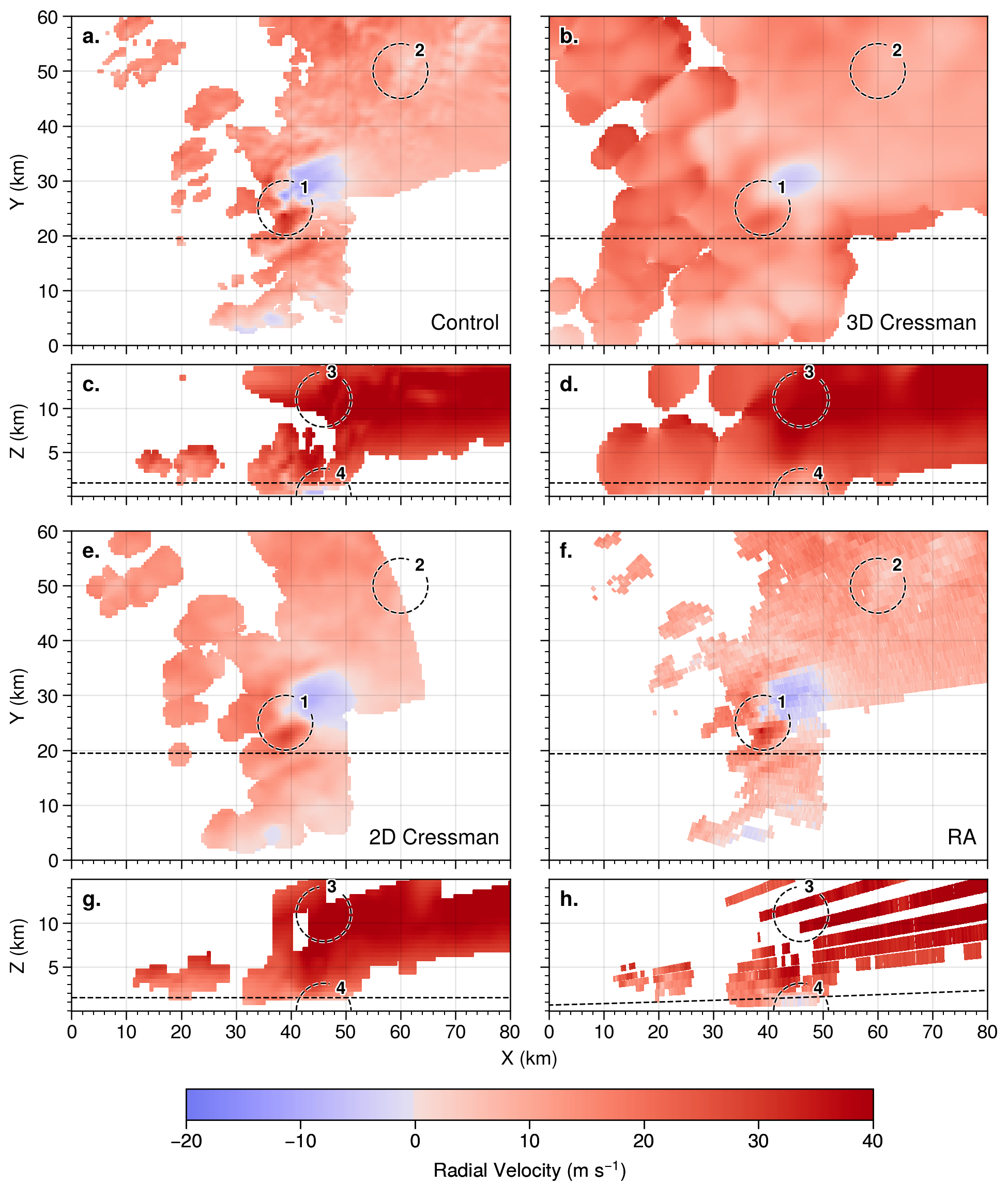}\\
 \caption{\label{obs} Simulated radial velocities for radar position 1 for each of the 3D wind retrieval methods in this study. (a), (b), (e) Horizontal cross sections at $Z$ = 1.5 km. (f) PPI from the second sweep ($\theta = 0.8^\circ$). (c), (d), (g) Vertical cross sections at $Y$ = 19.5 km. (h) “Pseudo” RHI from the 90$^\circ$ azimuth, where radar data are shaded inside the bounds of the 1$^\circ$ angular beamwidth of each ray. Straight dotted lines indicate the positions of the corresponding cross sections, and circular regions are highlighted for discussion in the text.}
\end{figure*}

In this section, we analyze the performance of the first of fifteen ensemble members pictured in Fig.\ \ref{setup} (\ie experiment 1 of 15). We aim to derive a qualitative understanding of the performance of each retrieval method through the analysis of this single experiment, before providing more quantitative ensemble statistics in Section \ref{results}\ref{ss:ensemble}. Firstly, Fig.\ \ref{obs} illustrates the various radial velocity inputs for the northwesternmost radar in experiment 1. The control field in Figs. \ref{obs}a and \ref{obs}c has not undergone any radar sampling, rather, it is a direct projection of the true Cartesian wind components in the radial direction. This setup renders it a ground truth description of the underlying radial velocities, limited spatially to regions with radar reflectivity $\geq$ 5 dBZ. 

Perhaps the most noteworthy signature in the control field is the pronounced velocity couplet in region 1 of Fig.\ \ref{obs}a, indicating the presence of a strong low-level mesocyclone. The shape and azimuthal shear values across the couplet are largely smoothed in the pre-gridded radial velocities in Figs. \ref{obs}b and \ref{obs}e. These effects are due to spatial aliasing in the radar gridding methods, which reliably filters observational noise, at the expense of high-frequency information required to accurately resolve such small-scale features (B22). We hypothesize that this type of smoothing will hamper the ability of pre-gridded methods to retrieve important dynamical features such as mesocyclone rotation in the resulting 3D wind retrievals. In contrast, the simulated radar data used in the RA method (Fig.\ \ref{obs}f) accurately resolves the velocity couplet at $\sim$60 km range in experiment one, albeit in the presence of notable speckle noise added during the radar simulation process. This level of observational noise is common throughout the simulated radar data (\eg in region 2), and it is not clear \textit{a priori} whether the benefits of finer data resolution will outweigh the added observational noise in the RA method. 

Another important consideration for each wind retrieval method is the spatial continuity; or equivalently, the size and shape of data voids, in the radial velocity data. The ``true'' extent of the simulated radar echoes is shown in the control experiment (with the minimum detectable signal defined at 5 dBZ). The 3D Cressman method artificially extends valid data past its true extent (\eg between regions 3 and 4 in Fig.\, \ref{obs}), effectively ``filling in'' many true data voids. It is not obvious whether this data extension is advantageous in 3D wind retrievals, as the added observations may actually assist in retrieving valid winds in data voids, contingent on the quality of the extrapolated radial wind values. Velocity observations in the RA method contain large gaps aloft due to the spacing between constant elevation radar sweeps (\eg region 3 in Fig.\ \ref{obs}h), and this may have serious implications for the accuracy of retrieved winds if the interpolation operator ($\mathcal{C}$) and the smoothing constraints ($J_s$) cannot effectively propagate radar information into these data voids. Similarly, the radial velocities in the 2D Cressman method are severely limited above the highest tilt, and below the lowest tilt (\eg region 2 in Fig.\ \ref{obs}e and region 4 in Fig.\ \ref{obs}g). A notable consequence of these spatial continuity considerations may be observed in region 4, where a signature indicating easterly inflow in the low levels is evident in the control observations (Fig.\ \ref{obs}c). This signature is largely absent in the pre-gridded methods, due to over-smoothing in the 3D Cressman observations, and data gaps below the lowest tilt in the 2D Cressman observations. Once again, we hypothesize that the pre-gridded retrievals will be adversely affected by the omission of this important dynamical information in the underlying radial velocity data.

\begin{figure*}
\centering
\includegraphics[width=0.95\textwidth]{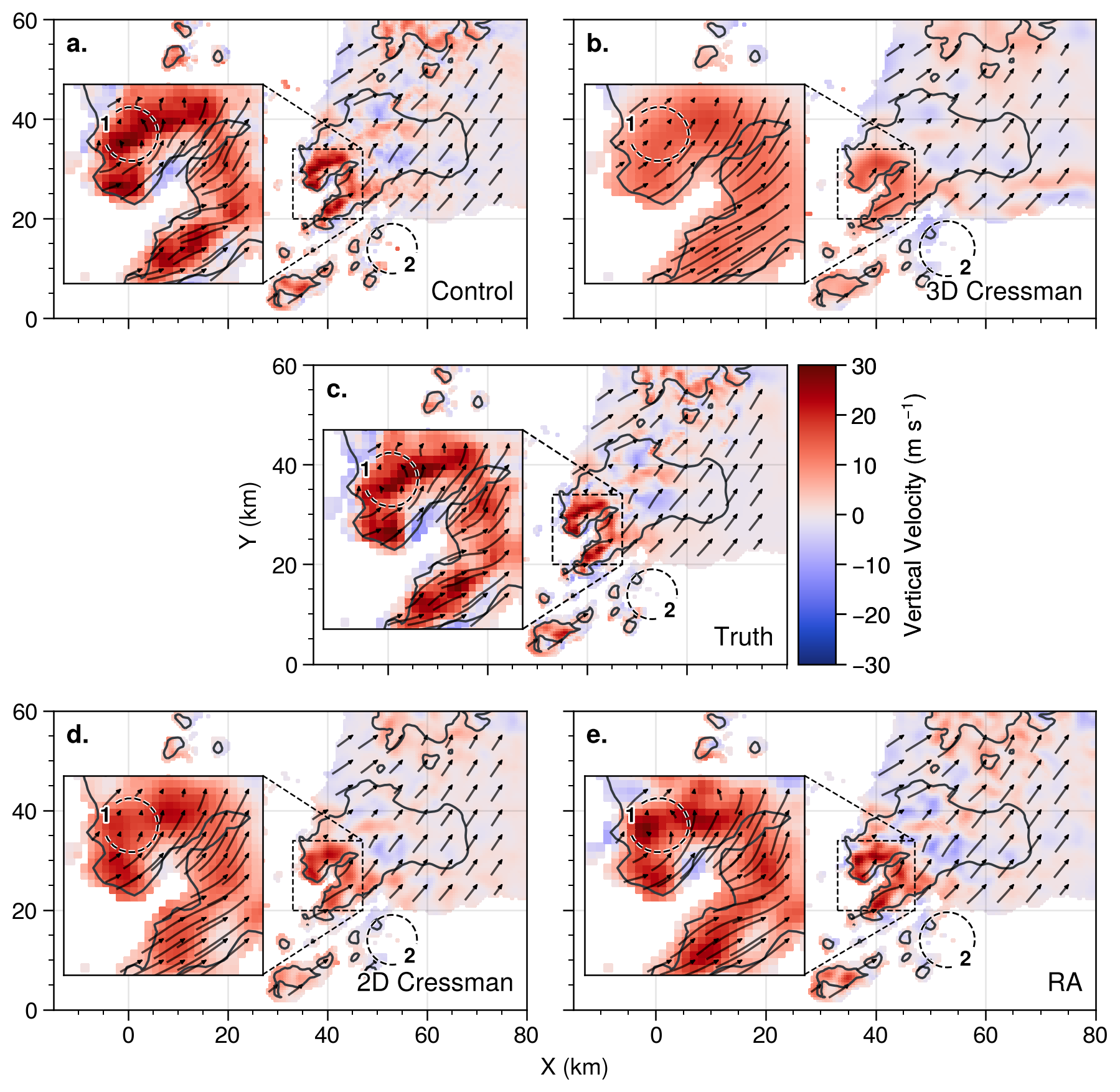}\\
 \caption{\label{horiz_plot}Horizontal cross sections at $Z$ = 5 km with a 50 dBZ reflectivity contour in gray, streamlines illustrating horizontal velocities (length proportional to horizontal wind magnitude), and shading indicating vertical velocities: (c) model ``truth'' winds, along with retrieved winds from the (a) control, (b) 3D Cressman, (d) 2D Cressman and (e) RA methods. Insets for each panel show a zoomed view of the analysis region discussed in the text (spanning X = 33 -- 47 km and Y = 20 -- 34 km).}
\end{figure*}

Horizontal cross sections of retrieved vertical velocity for each method are given in Fig.\ \ref{horiz_plot}. Visual comparisons of both shading colors and streamlines between Figs. \ref{horiz_plot}a and \ref{horiz_plot}c indicate that the control experiment is able to reproduce the true 3D wind field very accurately. One notable exception in the accuracy of the control experiment are small patches of spurious vertical velocity retrievals, such as those in region 2 of Fig.\ \ref{horiz_plot}a. These are due to boundary effects in data-sparse regions, and are an example of how errors are still present in 3D wind retrievals with ``perfect'' radial velocities. For these reasons, the accuracy of any 3D wind retrieval method simulated using practical radar scanning geometries should also be judged against such a control experiment, and not merely the ``true'' model winds. The reader is referred to Appendix B for further discussion on boundary effects in 3D wind retrievals. 

All three retrievals that ingest radar simulated radial velocities (either pre-gridded in 3D/2D Cressman methods, or directly in the RA method) are able to reproduce the broad-scale southwesterly horizontal flow, along with the approximate size and shape of the updraft within the inset region in Fig.\ \ref{horiz_plot} (hereafter referred to as the analysis region). The largest errors arise in the vertical velocity retrievals for the pre-gridded methods, which largely underestimate the true magnitude of the updraft velocities (by more than a factor of 2 in the 3D Cressman method), and are unable to retrieve the small downdraft features around the edges of the main updraft (blue shading, northwest and southeast of region 1 in Fig.\ \ref{horiz_plot}c). We attribute these poor vertical velocity retrievals to the errors introduced by pre-gridding the underlying radial velocity fields. More specifically, filtering and over-smoothing of the highest radial velocity magnitudes, along with the poorly resolved low-level convergence discussed in Fig.\ \ref{obs}, both contribute to the large vertical velocity errors noted in Fig.\ \ref{horiz_plot}. 

Unlike pre-gridded methods, the RA method does accurately resolve the amplitude and extent of the main updraft region (with maximum updraft velocities in Fig.\ \ref{horiz_plot}e approaching the true values evident in Fig.\ \ref{horiz_plot}c). Furthermore, the position of the aforementioned downdraft features are retrieved well in this method, while slightly underestimating maximum downdraft speeds. We attribute this success to the higher resolution of the input radial velocity data, which allows the technique to resolve the small-scale and highly dynamic circulation within the analysis region. Another example of the retrieval of important small-scale dynamic features is the cyclonic rotation within region 1 in Fig.\ \ref{horiz_plot}c. Streamlines in Fig.\ \ref{horiz_plot}e indicate this feature is retrieved well in the RA method, where the easterly component of the mesocyclone rotation is clearly visible against the prevailing westerly background flow. Both pre-gridded methods were unable to retrieve any easterly wind component within the mesocyclone region, which has important implications for identifying supercell thunderstorms, which cause a disproportionate amount of large hail and significant tornadoes \citep[\eg][]{Duda10, Smith12}.

\begin{figure*}[ht]
\centering
\includegraphics[width=1\textwidth]{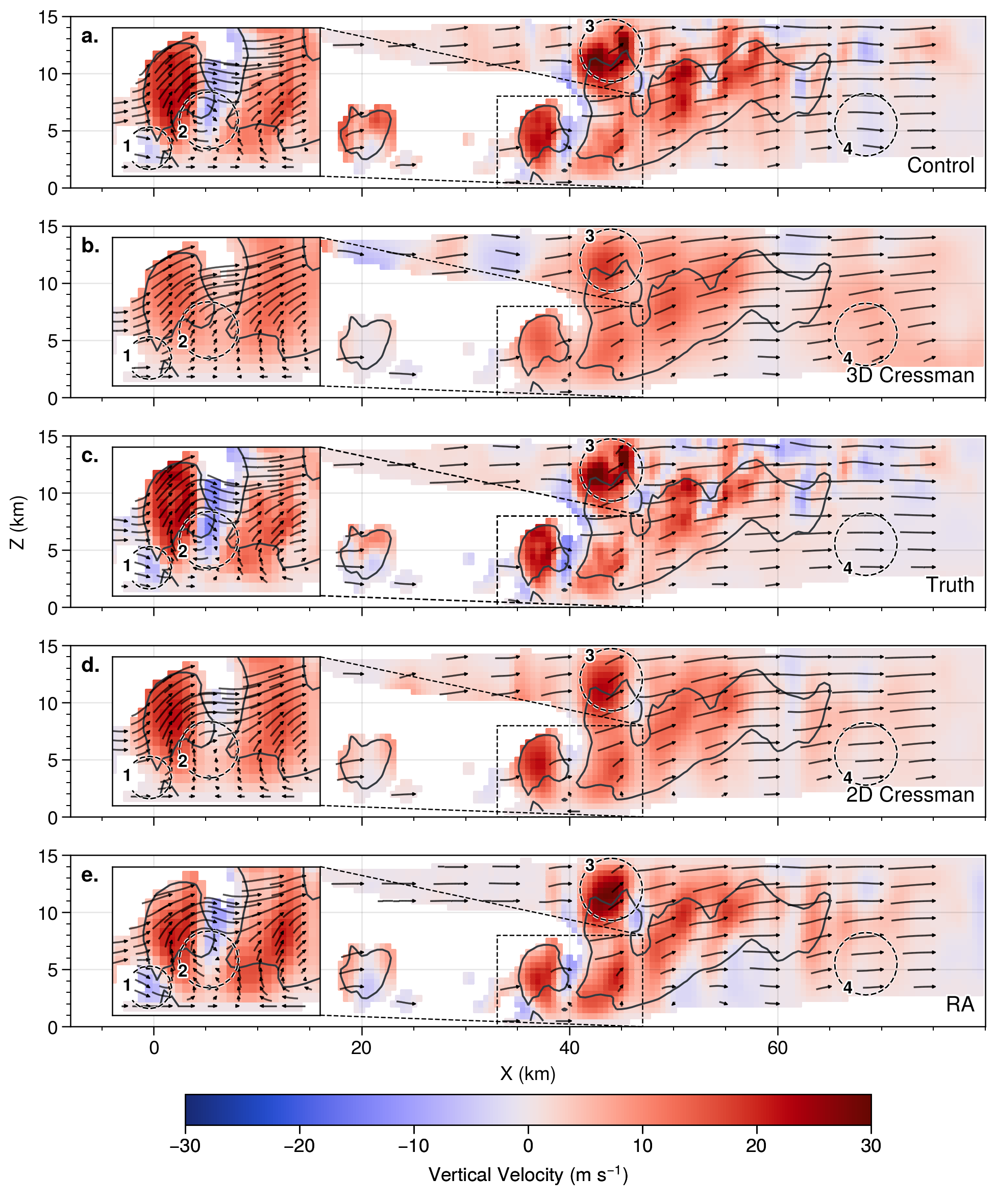}\\
 \caption{\label{vert_plot}As in Fig.\ \ref{horiz_plot}, but for a vertical cross section at Y = 30 km. The inset panel shows a zoomed view of a region spanning X = 33 -- 47 km and Z = 0.5 -- 8 km.}
\end{figure*}

Vertical cross sections of vertical velocities from experiment 1 in Fig.\ \ref{vert_plot} further illustrate the deficiencies of the pre-gridded retrieval techniques. We will highlight these deficiencies in the context of two important dynamic features present in the analysis inset in Fig.\ \ref{vert_plot}c: a rear flank downdraft (RFD) signature close to the surface within region 1, and an overturning circulation resulting from an updraft/downdraft couplet within region 2. Both pre-gridded methods were unable to retrieve any downdraft areas within the inset region, meaning both the RFD and overturning circulation are absent from the resulting wind fields. However, the RA was again able to retrieve these features accurately in terms of their shape and position, whilst slightly underestimating the true vertical velocity maximum values. The retrieval of these features has important implications for nowcasting meteorological hazards, such as identifying potentially strong straight line winds at the surface due to the RFD, and the existence of favorable hail growth trajectories due to the overturning circulation \citep[\eg][]{Kumjian21}. 

While the wind retrieval clearly benefits from the added resolution of radial velocity observations in the RA method, this method still displays considerable limitations when compared to the control retrieval. Firstly, consider the 3rd highlighted region in Fig.\ \ref{vert_plot}c, which contains two spatially proximal, but distinct regions of very high vertical velocities ($>$ 30 m s$^{-1}$). The control method in Fig.\ \ref{vert_plot}c accurately retrieves these separated updraft regions, meaning this scale of motion (these features, and most convective towers, are approximately 1--2 km in diameter) is practically resolvable on the 500 m isotropic analysis grid used in this study. The RA method is once again able to retrieve the magnitude of this updraft feature in region 3 much more accurately than the pre-gridded methods, however, the two updraft features are aliased into one large updraft region in Fig.\ \ref{vert_plot}e. This spatial aliasing is present throughout the upper levels ($>$10 km), and is a result of the sparsity of observed radar information at these altitudes (refer to Fig.\ \ref{obs}h). Clearly, the accuracy of the RA method is still fundamentally limited by the scanning geometry of the underlying radar observations, albeit to a lesser extent than the pre-gridded methods. Finally, we also note the poor performance of all three radar-derived retrieval methods in the forward flank of the storm (\eg region 4 in Fig.\ \ref{vert_plot}). Here, spurious vertical velocity signatures are present throughout the depth of the storm, and are coincident with significant data gaps due to clear air in the low levels.  We posit these errors are caused by a well-known problem in the wind retrieval literature, whereby the impermeability boundary condition ($w = 0$ at $Z = 0$) in the mass continuity constraint is poorly imposed in regions with data voids near the surface. Studies have shown these issues may be alleviated through the inclusion of a vertical vorticity constraint \citep{Shapiro09, Potvin12b}, which is not implemented here due to our sole focus on the effects of spatial interpolation.

\begin{figure*}
\centering
\includegraphics[width=\textwidth]{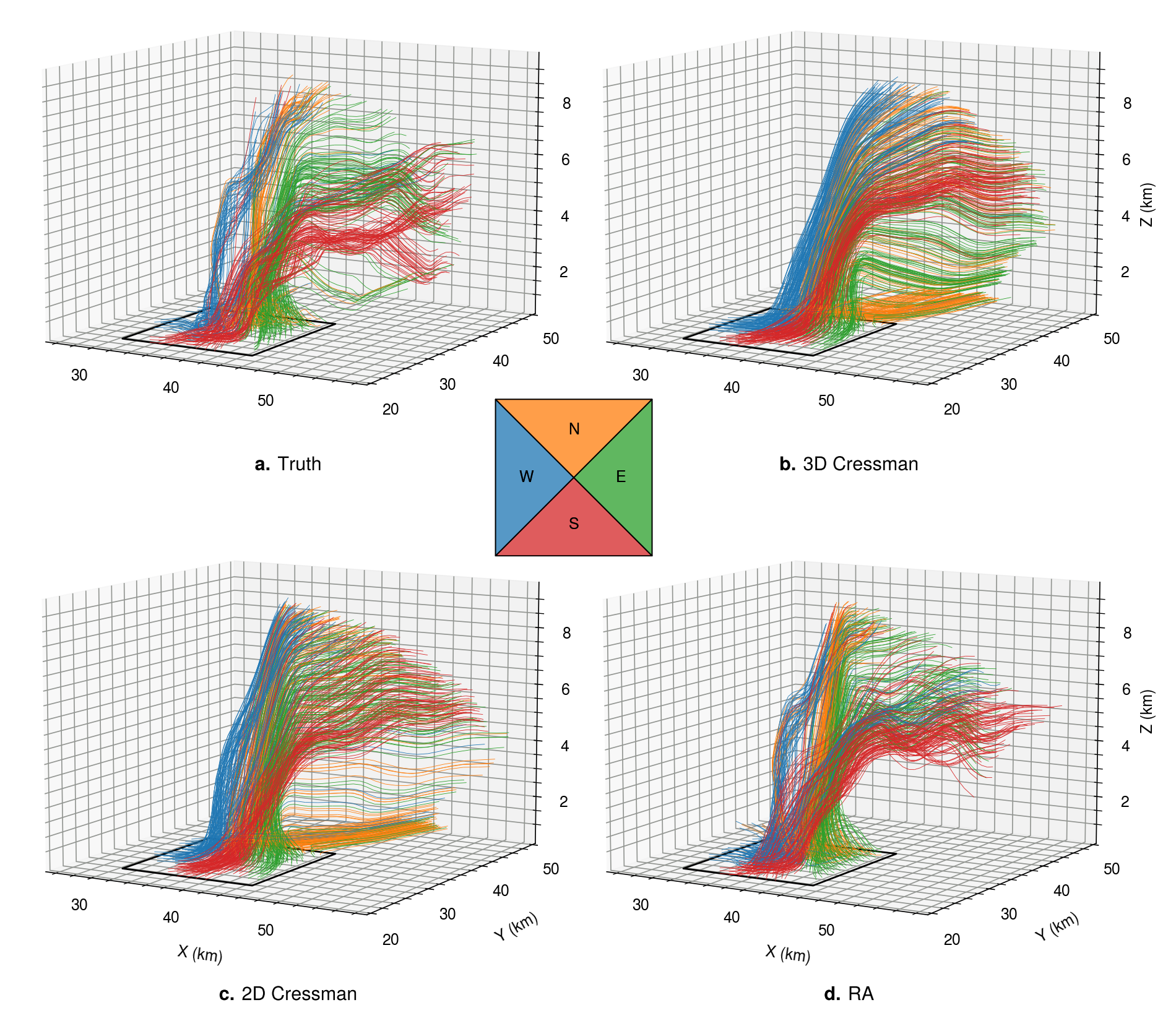}\\
 \caption{\label{3d_plot} Trajectory calculations for air parcels initiated once every 1 km in the horizontal dimensions within the analysis region (pictured in black), and 600, 700 and 800 m in altitude. Trajectories are modelled for (a) true model wind fields, and (b)-(d) retrieved wind fields, and colored according to their initial quadrant position (refer to the inset for details on quadrant coloring).}
\end{figure*}

The final qualitative assessment of the wind fields from experiment 1 is a comparison of simulated air parcel trajectories in Fig.\ \ref{3d_plot} (the control experiment is omitted due to its similarity with the model truth). Trajectories are initiated within the inflow of the storm at 600 m, 700 m and 800 m altitudes, at 1 km spacing within the horizontal bounds of the analysis box shown in the inset in Fig.\ \ref{horiz_plot}. Broadly speaking, trajectories calculated using the true wind fields in Fig.\ \ref{3d_plot}a follow two main paths: strong vertical ascent to the top of the domain ($>$8 km) within the updraft [mostly those initiated in the western (blue) and northern (orange) quadrants], or initial ascent followed by ejection into the forward flank of the storm in the mid-levels [4--8 km, mostly the eastern (green) and southern (red) quadrants]. The RA winds in Fig.\ \ref{3d_plot}d produce trajectories that are qualitatively similar to the true trajectories for all four quadrants. Most notably, the RA method reproduces the strong split between trajectories from the southern (blue) and western (red) trajectories, which is also observed in Fig.\ \ref{3d_plot}a. Many trajectories in the pre-gridded fields also exhibit very similar trajectories to those in Fig.\ \ref{3d_plot}a, however, there are also some notable differences. For example, the aforementioned split between the western and southern inflow regions is not observed in Figs. \ref{3d_plot}b and \ref{3d_plot}c. Rather, there exists a smooth transition between these pathways, resulting in a roughly continuous spatial distribution between the trajectories at the top of the storm and those ejected into the mid-levels. 

Another notable difference in the pre-gridded retrievals (\vs the radar assimilation retrievals) is the comparative smoothness, or uniformity, of the trajectories, especially in the forward flank. This is once again a result of filtering of important information in the underlying gridded radial velocities, resulting in over-smoothed retrieved wind fields. Lastly, perhaps the most notable difference in the pre-gridded fields is the spurious collection of low-level trajectories which initiate from the northern quadrant of the initiation region. The presence of these trajectories indicates a lack of convergence within the main updraft, which ultimately results in the aforementioned underestimation of updraft strength for these methods. The lack of convergence in the low-levels was hypothesized through the analysis of Fig.\ \ref{obs}, where the inflow signature was poorly resolved in the pre-gridded radial velocities. Fig.\ \ref{3d_plot} qualitatively confirms this assumption, and emphasizes the importance of ingesting accurate radial velocity information, especially in the low levels.

\begin{table*}[t]
\centering
\caption{\label{tab} Summary error statistics for the 15 member ensemble for each retrieval method. The ensemble mean and standard deviation (in parentheses) are shown for the total RMSE, RMSE within the analysis region (an.) and RMSE for individual wind components ($u$, $v$, $w$). Fractions Skill Scores (FSS) are also listed for updrafts (up, 95th percentile of the column maximum vertical winds) and downdrafts (down, 5th percentile of the column minimum vertical winds).}
\begin{tabular}{lrrrrrrr}
\toprule
{} &         RMSE &      RMSE an. &     RMSE $u$ &     RMSE $v$ &     RMSE $w$ &       FFS up &     FSS down \\
\midrule
Control     &  2.00 (0.05) &   2.68 (0.04) &  0.23 (0.13) &  0.22 (0.11) &  1.96 (0.03) &  0.95 (0.00) &  0.80 (0.02) \\
3D Cressman &  7.17 (0.16) &  11.61 (0.22) &  4.42 (0.28) &  4.25 (0.36) &  3.68 (0.16) &  0.70 (0.04) &  0.26 (0.04) \\
2D Cressman &  7.43 (0.37) &   8.90 (0.37) &  4.96 (0.45) &  4.28 (0.29) &  3.49 (0.14) &  0.76 (0.04) &  0.43 (0.05) \\
RA          &  6.22 (0.27) &   8.03 (0.61) &  3.77 (0.57) &  3.61 (0.52) &  3.30 (0.13) &  0.88 (0.01) &  0.63 (0.06) \\
\bottomrule
\end{tabular}
\end{table*}

\subsection{Ensemble Analysis}\label{ss:ensemble}

Table \ref{tab} presents a summary of the experimental results for each retrieval method, including ensemble mean and standard deviation values. The control experiment achieves excellent results across all radar geometries tested in this study. Considering the average wind magnitude for the experiment is $\sim$51 m s$^{-1}$, the ensemble average RMSE of 2.00 m s$^{-1}$ results in an average percentage error of just $\sim$4\% in wind magnitudes across the domain. A standard deviation of 0.05 m s$^{-1}$ for the 15 experiments also indicates the technique is not particularly sensitive to different radar positions or cross beam angles. As expected, the vast majority of error comes from the poorly observed vertical wind component, which has an average RMSE of 1.96 m s$^{-1}$, compared to the horizontal component RMSE's of 0.23 and 0.22 for $u$ and $v$, respectively. We also note the control experiment is able to retrieve winds with an average RMSE of 2.68 m s$^{-1}$ in the analysis region, which indicates that the ``perfect'' radial velocities on a 500 m analysis grid are capable of resolving the scales of motion very well in this highly dynamic region of the storm. Finally, we observe that Fractions Skill Scores (FSS) for updrafts outperform those for downdrafts in the control experiment (0.95 \vs 0.80, respectively), and note this finding is consistent across the three radar observed methods. This is likely a result of the proximity of downdraft regions to the top of the analysis domain ($z$ = 15 km) at this time step, where boundary errors are prevalent (compare downdrafts in Figs. \ref{vert_plot}a and \ref{vert_plot}c). 

As expected from the qualitative analysis in Section \ref{results}\ref{ss:case}, the quality of retrievals is considerably degraded when winds are retrieved with simulated radar radial velocities, resulting in a 300\% -- 350\% increase in error magnitudes. Interestingly, we did not observe reliable accuracy reductions for ensemble members with more distant radars (not shown), indicating the storm was adequately sampled by the volume coverage pattern at each of the three ranges used in this study (60, 70 and 80 km). Firstly, both pre-gridded methods score similarly in terms of overall RMSE scores (7.17 m s$^{-1}$ for 3D Cressman and 7.43 m s$^{-1}$ for 2D Cressman), however, ensemble standard deviations indicate that error scores are much more consistent in the former. This is easily explained by the superior error filtering qualities in the 3D Cressman analysis, which limits the amount of observational error propagating into the retrieved wind fields. While the smoothing properties of the 3D Cressman analysis are beneficial in the consistency of error scores, they also hinder the retrieval of strong winds in the analysis region, where the RMSE score is considerably higher at 11.61 m s$^{-1}$. The inability of the 3D Cressman method to retrieve winds in highly dynamic regions is further underscored by the poor updraft and downdraft FSS values (defined by the 5th and 95th percentile of column-maximum vertical velocities, $w\approx8.5$ and $w\approx-3.4$, respectively) relative to the control experiment (0.70 \vs 0.95 for updrafts, and 0.26 \vs 0.80 for downdrafts). These results ultimately illustrate the trade-off between the 3D Cressman and 2D Cressman gridding methods: the former results in more consistent retrievals with slightly lower total error measures, but degrades the retrieval in the most dynamic regions of the storm, which are perhaps the most important in terms of understanding storm dynamics and nowcasting potential hazards.

\begin{figure*}
\centering
\includegraphics[width=\textwidth]{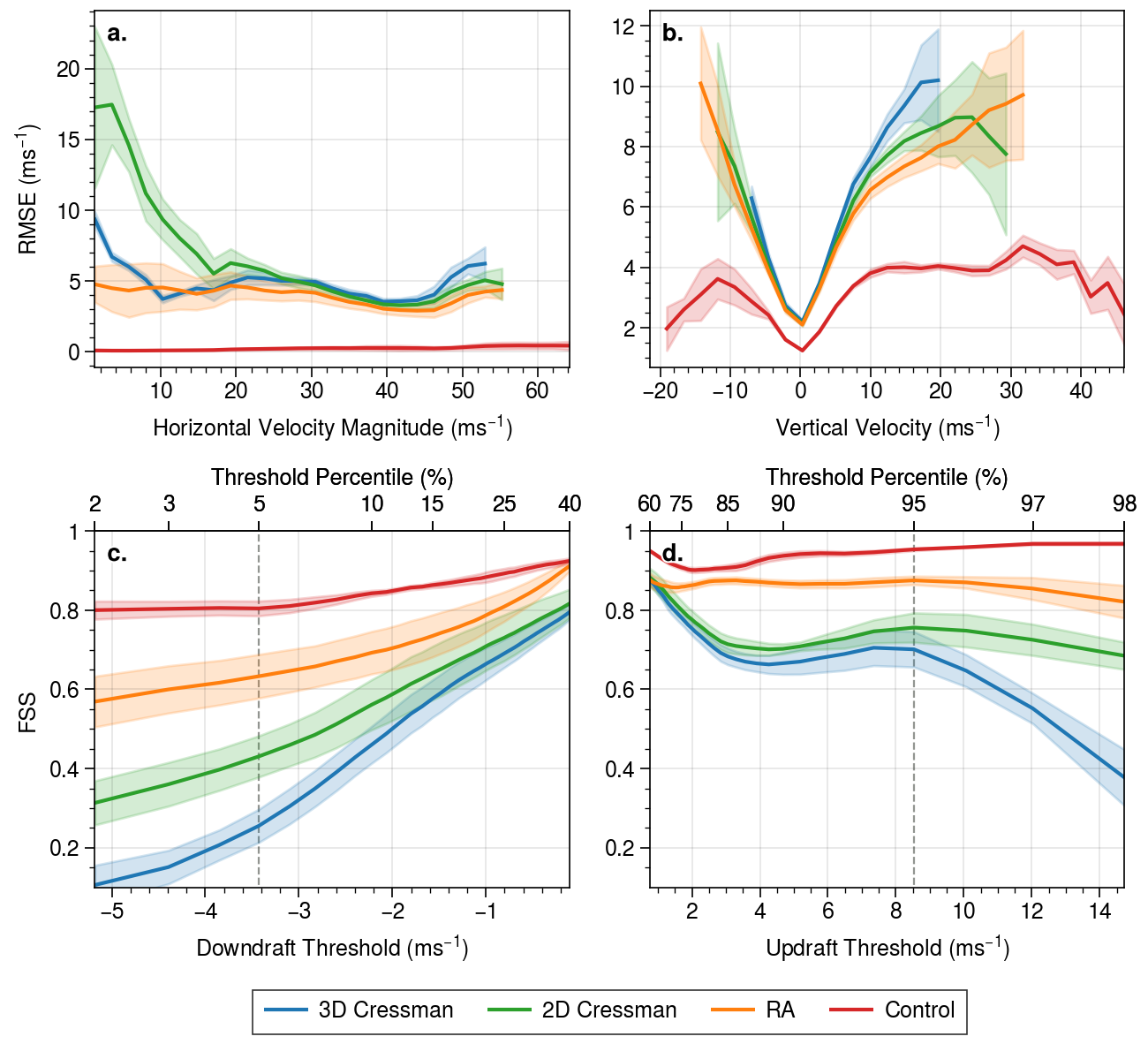}\\
 \caption{\label{magnitude} Root mean square errors (RMSE) across a range of magnitudes for various quantities and retrieval methods. All values are classified into 30 bins according to their magnitude between the maximum and minimum of the true values for each field. The RMSE is calculated within each bin, and the ensemble mean error is plotted with a solid line. One standard deviation of the ensemble values is shown above and below the mean to indicate the ensemble spread. A line is plotted at each magnitude only if all 15 ensemble members have at least one valid value in the bin, and thresholds used to compute FSS values are indicated by gray dotted lines in panels (c)-(d).}
\end{figure*}

The average RMSE score of 6.22 m s$^{-1}$ in the RA method supports the aforementioned qualitative improvements over the pre-gridded methods. The use of this technique results in an average reduction in errors of 18\% and 22\% compared with the 3D Cressman and 2D Cressman methods, respectively, relative to the control experiment. These domain-wide improvements relative to the pre-gridded methods are also reflected in RMSEs in the analysis region (8.01 m s$^{-1}$), and for individual wind components over the entire domain ($\sim$3.5 m s$^{-1}$ errors for $u$, $v$ and $w$). Interestingly, while the average vertical velocity error is slightly lower in the RA method, the majority of RMSE improvements come from reducing errors in the horizontal components, which are  much lower than in the pre-gridded retrievals. This belies the qualitative observations made in Figs. \ref{horiz_plot} and \ref{vert_plot}, which showed considerable improvements in RA vertical winds, especially within the analysis region. An explanation for this seeming discrepancy is that high-amplitude fields are penalized more heavily by pixel-to-pixel accuracy measures such as RMSE, as slight spatial offsets in wind fields result in higher error scores compared to smoother fields such as the 3D Cressman method. This effect also accounts for the lower RMSE values observed in the 3D Cressman method compared to the 2D Cressman method, despite the considerable over-smoothing noted in the former in Section \ref{results}\ref{ss:case}. We further this hypothesis by noting that the FSSs for updraft and downdraft regions are considerably higher for the RA method. The updraft and downdraft FSSs in the RA method are only 7\% and 21\% lower, respectively, than the control experiment, while the 3D Cressman and 2D Cressman methods show larger a reduction of 20\% and 46\%, and 26\% and 68\%, respectively. These statistics indicate that whilst the RA method may not achieve a significantly lower RMSE for vertical velocities, it does retrieve regions of high vertical velocity much more reliably than the pre-gridded methods. 

After presenting the mean RMSE and FSS values for the entire domain in Table \ref{tab}, we proceed to examine the variation of these statistics with the magnitude of the underlying variables in Fig.\ \ref{magnitude}. Firstly, Fig.\ \ref{magnitude}a confirms that horizontal velocities are retrieved very well with the ``perfect'' observations in the control experiment ($<$0.5 m s$^{-1}$ errors across the entire range of magnitudes). For the other retrieval methods, the sampling effects of the simulated radar observations considerably degrade the accuracy of horizontal winds, with comparatively large RMSEs ($>$3 m s$^{-1}$). The lowest horizontal velocity errors actually arise in regions with large velocity magnitudes ($\sim$45 m s$^{-1}$), which correspond to the mean environmental flow in the upper levels of the storm (8--12 km, refer to Fig.\ \ref{vert_plot}). Horizontal velocity errors are lower here due to the uniformity of the flow, with limited storm-induced perturbations and smaller vertical velocities. In contrast, regions with low horizontal velocities ($<$10 m s$^{-1}$) occur only very close to the surface, or in strongly dynamic regions of the storm, which oppose the broad-scale environmental flow (such as within the mesocyclone, refer to Fig.\ \ref{horiz_plot}). Both of these regions are poorly resolved in the pre-gridded retrievals, leading to large errors at low velocity magnitudes (especially for the 2D Cressman method, see Fig.\ \ref{magnitude}a). The 3D Cressman method also predictably shows larger errors for the strongest horizontal velocities, as the over-smoothed radial velocities cannot reproduce the highest magnitude winds. Generally speaking, the RA method produces the lowest horizontal velocity errors and has the most consistent accuracy across the range of horizontal velocity magnitudes. 

\begin{figure*}
\centering
\includegraphics[width=.92\textwidth]{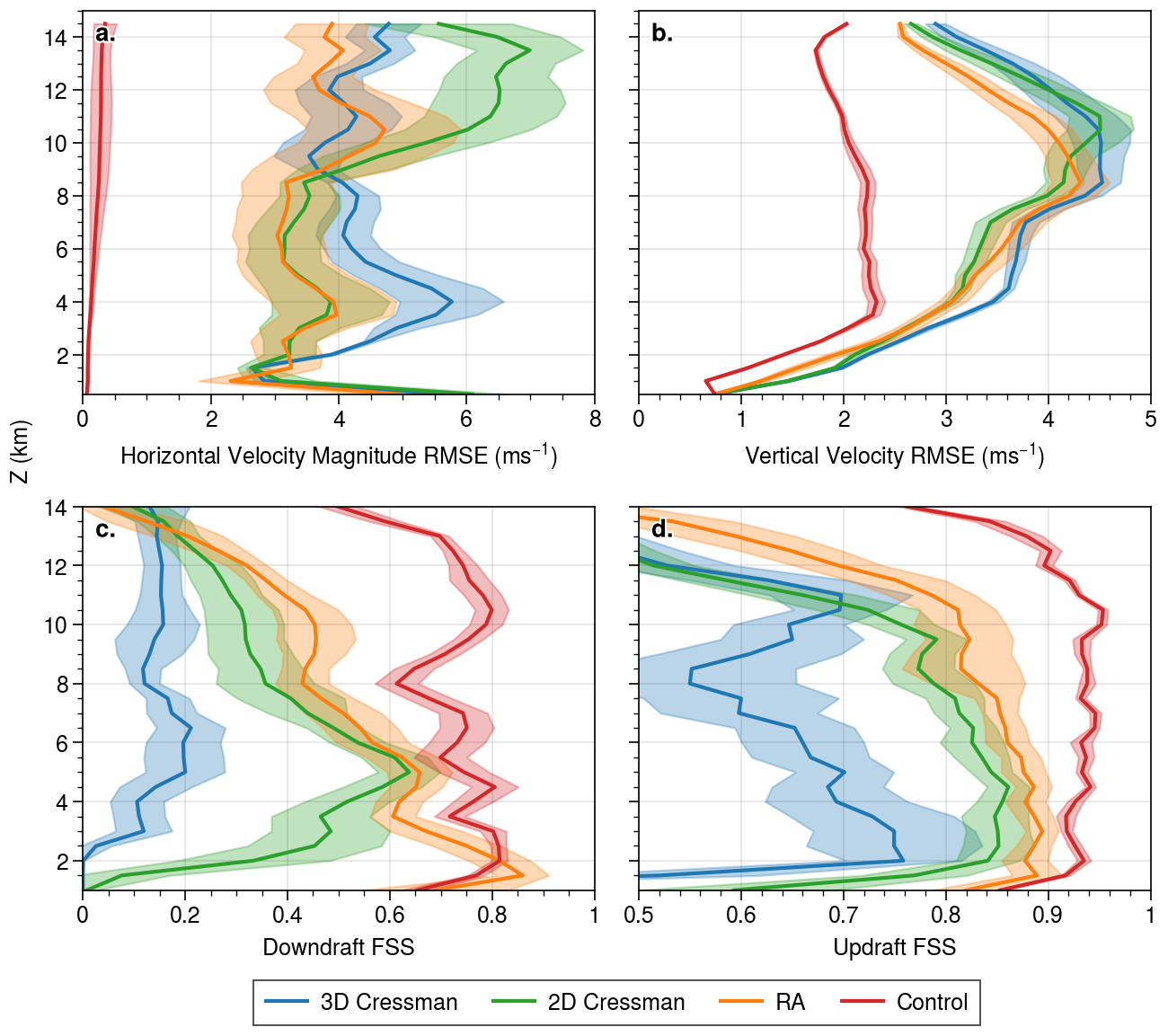}\\
 \caption{\label{vertical_stat} Vertical profiles of various error scores for each retrieval method. The quantities are grouped and averaged at each altitude level, and the ensemble mean and standard deviation are then calculated and plotted, as described in Fig.\ \ref{magnitude}.}
\end{figure*}

The difference in error magnitudes between the control experiment and the other retrieval methods is less pronounced for vertical velocities (compare the red lines in Figs. \ref{magnitude}a and \ref{magnitude}b). This reflects the comparative difficulty of retrieving accurate vertical velocities with low-elevation radial velocities, even with ``perfect'' radar observations. Errors are lowest for all methods in areas with small vertical velocities ($\sim$0 m s$^{-1}$), and generally increase at higher velocity magnitudes. As in the horizontal velocities, 3D Cressman retrievals most poorly resolve the highest wind magnitudes due to over-smoothing of important high-frequency information, and the RA method produces the lowest errors across most vertical velocity magnitudes. The RA method is also able to retrieve considerably larger vertical velocity magnitudes (both updrafts and downdrafts) relative to the pre-gridded techniques, however, these retrieved maximum values are still considerably underestimated. We also examine how FSS values vary with respect to the thresholds used to delineate the updraft and downdraft regions. We test thresholds between the 60th and 98th percentile of the column-maximum vertical velocity field to compute updraft FSSs, and between the 2nd and 40th percentile of the equivalent column-minimum field for downdrafts. Figs. \ref{magnitude}c and \ref{magnitude}d show the qualitative FSS results presented in Table \ref{tab} (which used the 5th and 95th percentile values) are not sensitive to this threshold choice, as the RA method consistently outperforms the pre-gridded methods over the range of thresholds. This is particularly evident for the updraft and downdraft thresholds at the 98th and 2nd percentiles, confirming the RA method is better suited to retrieving regions with the most intense vertical velocities.
These ensemble findings concur with the qualitative observations of vertical velocities made based on Figs. \ref{horiz_plot} and \ref{vert_plot} in Section \ref{results}\ref{ss:case}. 

We also analyze how each of the diagnostics presented in Table \ref{tab} vary with altitude in Fig.\ \ref{vertical_stat}. Consistent with Fig.\ \ref{magnitude}, the control experiment exhibits very small horizontal velocity errors throughout the depth of the domain, but most notably at the lowest grid point (500 m altitude). This indicates that the impermeability condition imposed at the lowest grid point, as opposed to the ground surface, does not adversely affect the retrieval in this case \citep[a more sophisticated implementation considering the surface elevation should be used in areas with complex terrain;][]{Chong00, Liou12}. All three radar sampled methods exhibit large horizontal velocity errors ($>$5 m s$^{-1}$) at the surface due to the missing data below the lowest radar tilt elevation (0.5$^\circ$), especially the 2D Cressman method (refer to Fig.\ \ref{obs}g). Recall that Table \ref{tab} showed the majority of RMSE reductions in the RA method were attained through improvements in the horizontal velocity components. Fig.\ \ref{vertical_stat}a illustrates the origin of these improvements. The RA method produces roughly consistent RMSE scores throughout the depth of the storm ($\sim$4 m s$^{-1}$), however, both the pre-gridded methods show considerable deficiencies at different height levels. Firstly, the 3D Cressman method produces large horizontal velocity errors in the mid-levels of the storm ($\sim$4 km) due to a combination of not resolving the small-scale dynamics such as the mesocyclone within the core of the storm (\eg region 1 in Fig.\ \ref{horiz_plot}b), and boundary effects due to missing low level data within the forward flank of the storm (\eg region 4 in Fig.\ \ref{vert_plot}b, also refer to Appendix B). Secondly, the 2D Cressman method exhibits large errors in the upper-levels ($>$ 10 km), where the linear interpolation scheme used to fill large data gaps between PPI scans propagates observational noise and creates first-order discontinuities in the underlying radial velocity data \citep[][B22]{Trapp00, Askelson00}. The ability of the RA method to mitigate these error sources results in overall lower RMSE scores. 

Errors in vertical velocities for all four methods (including the control) increase steadily to a height of roughly 4 km, coinciding with the lower portion of the strong ($>$30 m s$^{-1}$) updraft in the mid-levels (refer to Fig.\ \ref{vert_plot}c). Above 4 km, the errors in the control experiment stay roughly constant at around 2 m s$^{-1}$, whereas the radar sampled methods are significantly degraded above 8 km. We attribute this poor performance to the operational radar scanning pattern used in our experiments, which results in large elevation gaps that fundamentally limit the accuracy achievable in the upper levels. As noted in Table \ref{tab}, the average improvement in vertical velocities gained by the RA method is modest throughout the depth of the storm, with 2D Cressman retrievals actually producing the lowest mean RMSE between 5--9 km in altitude. Again, we note that this result contrasts with the substantial qualitative improvements in vertical velocity observed for the RA method in \ref{horiz_plot} and \ref{vert_plot}, and speculate that the discrepancy may be due to the RMSE scores being biased in favor of pixel-to-pixel comparisons of smoother analysis fields. To verify this assertion, we recompute updraft and downdraft FSSs for the 5th and 95th percentile thresholds at each horizontal slice of the analysis grid. Figs. \ref{vertical_stat}c and \ref{vertical_stat}d show that the RA method exhibits modest improvements in FSS compared to the 2D Cressman method between 5--9 km in altitude, despite having slightly higher RMSE scores over the same altitude range. This may indicate a slight spatial displacement in vertical velocities, or that errors from vertical velocities below the FSS thresholds contribute significantly to the RMSE in this region. Furthermore, Fig.\ \ref{vertical_stat}d illustrates particularly low FSS values and high ensemble spread for the 3D Cressman method, further emphasizing its inability to accurately capture highly dynamic regions of the storm. Overall, Fig.\ \ref{vertical_stat} shows that the RA method produces lower error scores, and more reliably retrieves updraft and downdraft regions throughout the depth of the storm. 

\begin{figure*}
\centering
\includegraphics[width=.92\textwidth]{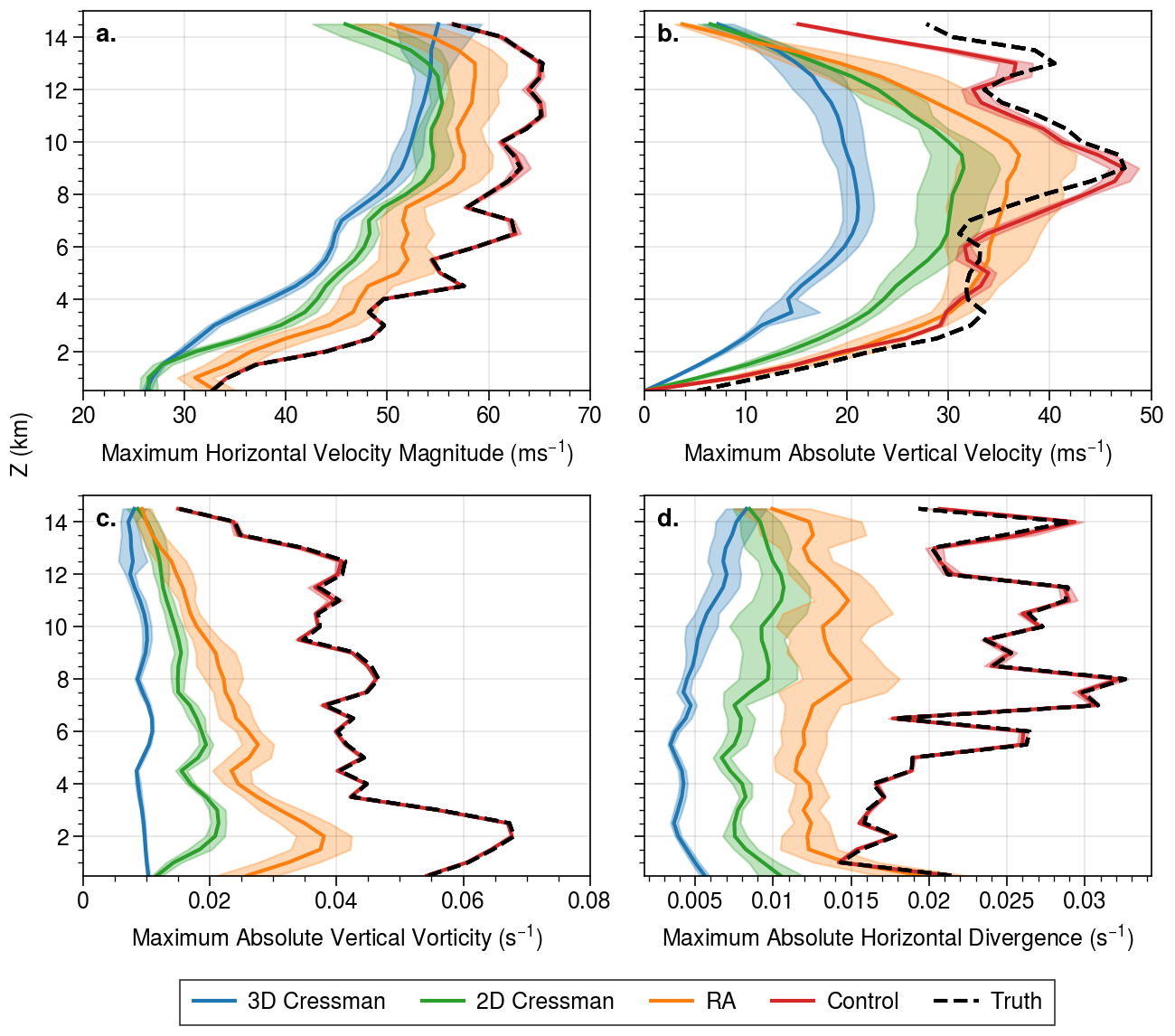}\\
 \caption{\label{vertical_mag} As in Fig.\ \ref{vertical_stat}, except for maximum values of various quantities. The `true' model maximum values for these quantities are also shown in black dashed lines.}
\end{figure*}

While the error statistics presented thus far are useful for understanding the error distribution over the entire domain (Table \ref{tab}), for various magnitudes (Fig.\ \ref{magnitude}), or over a range of altitudes (Fig.\ \ref{vertical_stat}), some analysts may prioritize accurately identifying the maximum values of certain quantities, such as the maximum horizontal velocity in the low-levels for forecasting severe wind hazards, or the maximum updraft velocity for diagnosing storm severity or verifying/informing convection parameterizations in numerical weather prediction models. These maximum value statistics are provided in Fig.\ \ref{vertical_mag}. As expected, the pre-gridded retrieval methods considerably underestimate the true horizontal and vertical velocity maxima, due to over-smoothing of important radial velocity information. Encouragingly, the RA method is able to accurately reproduce maximum vertical velocity values below 8 km, before considerably underestimating those farther aloft (at least partly due to the aforementioned observational gaps at high altitude). However, the poor performance of the control experiment at the domain top and bottom in Fig \ref{vertical_mag}b also indicates some of these errors may be partly attributed to the ill-posed vertical boundary conditions in the mass continuity equation.

Finally, the over-smoothing in pre-gridded methods is especially apparent in the derived maximum vorticity and divergence fields in Figs. \ref{vertical_mag}c and \ref{vertical_mag}d, and the qualitative effects of underestimating these dynamic quantities were observed in Figs. \ref{3d_plot}b and \ref{3d_plot}c, where simulated trajectories appeared unrealistically uniform or laminar. The RA method considerably outperforms these methods, while also underestimating the true vorticity and divergence maxima throughout the depth of the storm. The most notable improvements gained by the RA method occur in the lower levels in Fig.\ \ref{vert_plot}f, where the maximum convergence near the surface is retrieved accurately. This confirms the qualitative findings from the simulated radial velocity data in Fig.\ \ref{obs} and simulated trajectories in Fig.\ \ref{3d_plot}, which both suggested the superior characterization of low-level convergence and inflow in the RA method. The accurate characterization of low-level storm inflow is crucial for resolving storm dynamics \citep{Coffer22}, and this factor likely contributes to the qualitative and quantitative improvements observed from the RA method. 

\subsection{Real Data Case Study}\label{case}

\begin{figure}[t]
\centering
\includegraphics[width=0.5\textwidth]{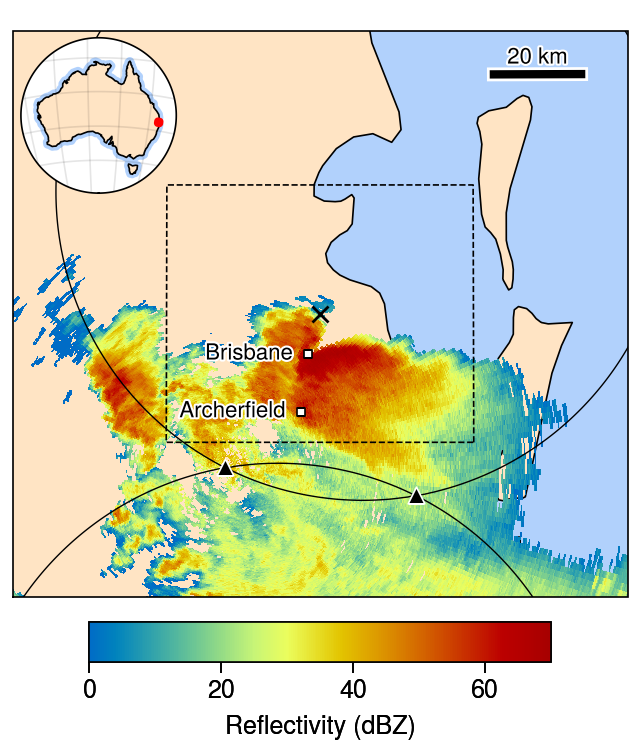}\\
 \caption{\label{domain} Experimental setup for the Brisbane hailstorm case study, indicating the center and extent of the analysis domain with a cross and dotted lines, respectively. Radars used in the analysis (triangles) and dual-Doppler lobes (solid lines) are shown alongside radar reflectivity from the 3.7$^{\circ}$ tilt at 0642 UTC. }
\end{figure}

In order to investigate the applicability of the OSSE findings to real cases, we now introduce a supercell case study from November 27, 2014 in Brisbane, Australia. This case was selected as an ideal candidate for verifying the ability to retrieve strong vertical motions, which are known to have occurred based on prior studies \citep[\eg][]{Parackal15, Soderholm17b}. The storm tracked northward through the Brisbane metropolitan area between 0200--0700 UTC, resulting in over AUD 1.5 billion in insured losses due to giant hail ($\sim$70 mm), severe straight-line winds, and localized flooding \citep{ICA2017}. The storm was observed by two S-band, Doppler radars: the CP2 research radar and the operational Mt.\ Stapylton radar (hereafter MS), located to the southwest and southeast of Brisbane, respectively. Fig.\ \ref{domain} shows the analysis domain for our dual-Doppler 3D wind retrieval, centered on the leading edge of the storm at (-27.4$^{\circ}$S, 153.05$^{\circ}$E), situated almost entirely within the dual-Doppler lobes ($\sim$40 km between radars). Quality controlled radar volumes are sourced from the Australian Unified Radar Archive \citep{Level1b}, containing nine and fourteen constant elevation sweeps for CP2 and MS, respectively\footnote{Exact elevations are $\theta$ = (0.9, 1.7, 2.4, 3.2, 4.7, 6.5, 9.1, 12.8, 17.8)$^{\circ}$ for CP2, and $\theta$ = (0.5, 0.9, 1.3, 1.8, 2.4, 3.1, 4.2, 5.6, 7.4, 10.0, 13.3, 17.9, 23.9, 32.0)$^{\circ}$ for MS. Half-power beamwidths/range gates are 0.93$^{\circ}$/150 m  and 1.0$^{\circ}$/250 m for CP2 and MS, respectively.}. 

In order to process real data, two minor methodological adjustments are made to the partially idealized OSSE retrieval methodology. First, we apply a reflectivity-based correction to account for the effects of hydrometeor terminal velocities in radial velocity measurements. We utilize the terminal velocity correction from the PyDDA package \citep{Jackson20}, which varies from pure rain \citep{Joss70} to hail \citep{Conway93} based on reflectivity thresholds. Second, we employ a simple horizontal advection correction to account for storm motion during the finite sampling period of radar volumes \citep[\eg][]{Shapiro09}. The average horizontal advection velocity was calculated by taking the mean optical flow velocity \citep[PySTEPS Lucas--Kanade implementation,][]{Pulkkinen19} for constant-altitude reflectivity slices between 1--5 km. Reflectivity grids were interpolated using the variational method outlined recently by \citet{Brook22}. Observations from both radars were then shifted to the mid-point of the radar scanning time (06:44 UTC) according to this average advection velocity prior to gridding or ingestion into the RA retrieval method. The analysis grid spans 60 and 50 km in the $x$ and $y$ axes, respectively, with a 500 m grid spacing. The grid is also spaced at 500 m increments in the $z$ axis, but extends from 200--15200 m ASL. 

\begin{figure*}
\centering
\includegraphics[width=0.99\textwidth]{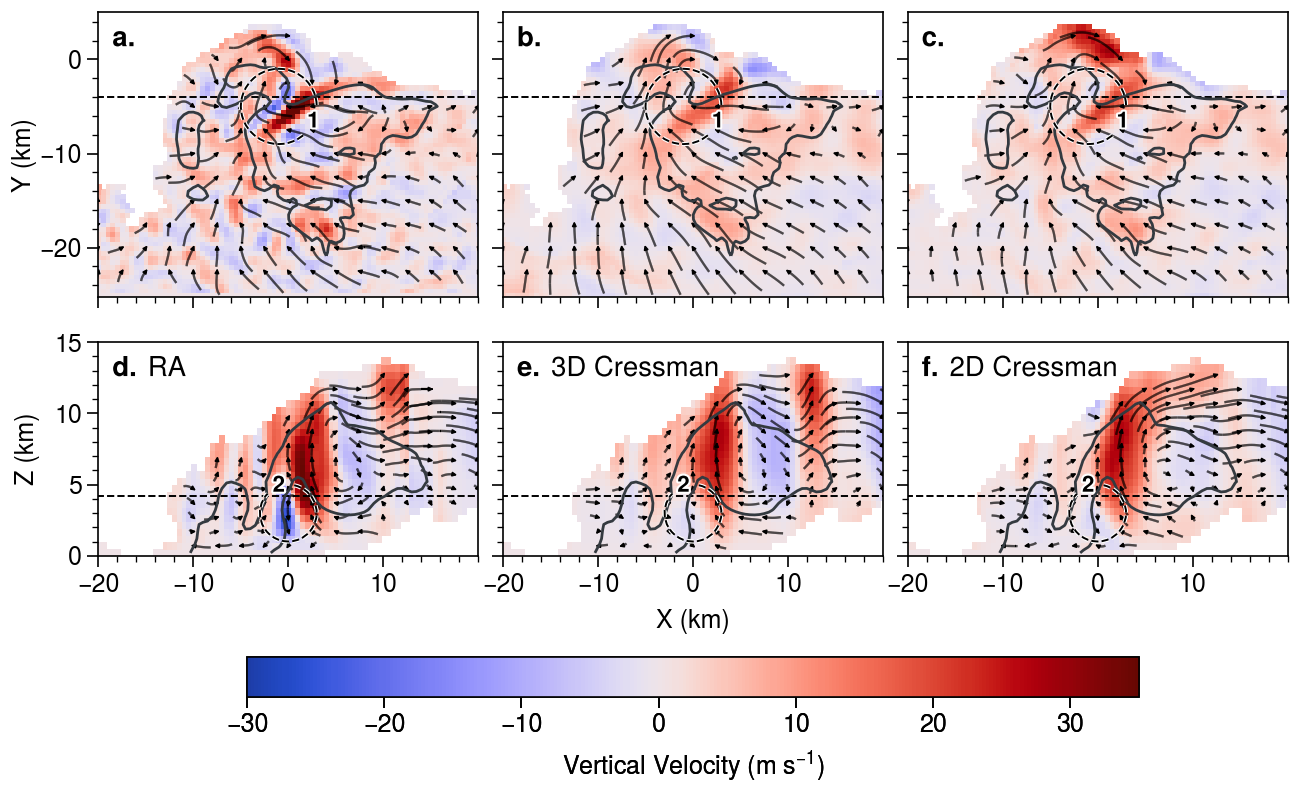}\\
 \caption{\label{brisbane} Retrieved winds for the 2014 Brisbane hailstorm for each of the methods discussed in this study. The 50 dBZ reflectivity contour is shown in gray, vertical velocities are shaded and streamlines illustrate along-plane velocities. (a)--(c) Horizontal cross sections at $Z$ = 4 km, and (d)--(f) vertical cross sections at $Y$ = -4 km. Straight dotted lines indicate the positions of the corresponding cross sections, and circular regions are highlighted for discussion in the text.}
\end{figure*}

Fig.\ \ref{brisbane} presents retrieved wind fields for the 2014 Brisbane hailstorm for each of the retrieval strategies discussed in this study. Horizontal streamlines in Figs.\ \ref{brisbane}a--c are broadly similar across each method, resolving the southeasterly change to the rear of the storm, and the strong mesocyclone circulation around highlight region 1 \citep[as in][]{Soderholm17b}. Similar to the OSSE experiments in Section \ref{results}\ref{ss:case}, horizontal and vertical velocities appear visually smoother in the pre-gridded retrievals (Figs.\ \ref{brisbane}b and \ref{brisbane}c), whereas small-scale structures in the vertical velocity fields, especially in the southwest of the domain, are resolved in the RA method (Fig.\ \ref{brisbane}a). The largest vertical velocity differences between methods occur to the north of region 1 in this case. All three methods capture an updraft at the leading edge of the storm, likely a result of dynamic lifting ahead of the advancing gust front. However, the leading-edge updraft appears spuriously large and intense for 2D Cressman method when compared to the main rotating updraft in region 1. This is likely a result of boundary effects caused by data scarcity at the storm's leading edge in the 2D Cressman method (refer to Appendix B), which under-represents the spatial extent of radar data, particularly in the low-levels (\eg region 4 in Fig.\ \ref{obs}g). 

The overall position and southwest--northeast orientation of the updraft at $Z=4$ km in highlight region 1 is retrieved by all retrieval methods. The shape of the inflow notch (indicated by the reflectivity contour in region 1) aligns well with the mesocyclone position and updraft orientation. Maximum updraft velocities for the RA method in Fig.\ \ref{brisbane}a are much larger than that in the pre-gridded retrievals. Quantitatively, the maximum updraft speed retrieved by the RA method was 52.5 m s$^{-1}$, compared to 30.8 and 30.3 m s$^{-1}$ for the 3D and 2D Cressman methods, respectively. While the veracity of the increased updraft velocities in the RA method cannot be directly confirmed in this real data experiment, the retrieved maximum value of $\sim$50 m s$^{-1}$ is closer to the mean values found in supercell simulations \citep{Peters20}, and the OSSE experiments suggest it is likely a more accurate estimate of the true updraft speed. Further qualitative evidence of considerable updraft strength for this case is given by the large, elevated hail core (>65 dBZ, $\sim$0 Z$_{\text{DR}}$), and bounded weak echo region well-above the freezing level \citep{Soderholm17b, Brook21}.  

Figs.\ \ref{brisbane}d--f corroborate these findings for a vertical cross section through the main updraft. The pre-gridded retrievals adequately estimate the position and shape of the updraft (as evidenced by the position of the bounded weak echo region in region 2 for Figs.\ \ref{brisbane}d--f), but underestimate vertical velocities relative to the RA method. Notably, the strong rear flank downdraft (RFD) signature shown in region 2 for the RA method is also entirely absent in the pre-gridded retrievals. This outflow feature is well documented for this event \citep{Parackal15, Soderholm17b, Brook21}, resulting in severe wind gusts and strong (>10 km) hail advection toward the west. The RA method's ability to resolve the RFD also leads to stronger divergence at lowest grid level, resulting in 40.2 m s$^{-1}$ maximum horizontal winds in the outflow region (west of region 2 in Figs.\ \ref{brisbane}d--f), compared with 24.0 and 30.3 m s$^{-1}$ for the 3D and 2D Cressman methods, respectively. By comparison, a 39.2 m s$^{-1}$ wind gust was measured roughly 10 minutes prior to the analysis time at Archerfield Airport (refer to Fig.\ \ref{domain}), along with widespread $\sim$20 m s$^{-1}$ gusts throughout Brisbane's western suburbs \citep{Parackal15, Soderholm17b}. While the difficulties associated with comparing retrieved winds aloft to wind measurements at the surface preclude a more quantitative comparison\footnote{Retrieved winds at the lowest grid level ($\sim$170 m AGL) are calculated using a free slip, impermeable, constant elevation boundary condition, and do not resolve the small-scale interactions (\ie due to topography and surface roughness) that strongly influence wind gust measurements at the surface.}, the strength and position of these surface measurements provide qualitative support for the RFD signature retrieved by the RA method. 

Overall, the results for the Brisbane supercell case study support the findings of the OSSEs, suggesting that the advances made in this study may be realized in practice. However, the practical implementation of the RA method also warrants a brief discussion on the relative computational efficiency of each retrieval method. The introduction of an additional interpolation step in the observational operator for the RA method (\ie $\mathcal{P}\mathcal{C}$ rather than $\mathcal{P}_g$, refer to Section \ref{methods}\ref{retrieval_method}) roughly doubles the computational cost of the operator relative to the pre-gridded methods. The observational constraint is only one of four otherwise identical constraints in the cost function (Equation \ref{ret_cost}), however, it must be performed twice at each iteration of the optimization procedure (forward and adjoint). In our implementation, the overall execution time of the RA method is thus $\sim$2.3 times slower than the pre-gridded methods for this case study (15.5 \vs 6.8 seconds, respectively). However, when considering the additional costs involved in pre-gridding the observations (4.4 and 8.0 seconds for 3D and 2D Cressman, respectively), the RA method introduces a relatively modest computational overhead.

\section{Summary and Future Directions} \label{summary}

The use of 3D wind retrievals for applications such as storm-scale dynamics research, constraining convection parameterizations, and nowcasting wind hazards is set to increase in the future, with open-source implementations such as PyDDA \citep[][]{Jackson19} increasing in popularity within the radar science community. Despite previous studies noting that spatial interpolation is a significant source of error in these analyses, there is currently no direct comparison or practical recommendations for spatial interpolation methodologies. In this study, we aimed to fill these knowledge gaps for severe convective storms by analyzing supercell 3D wind retrievals for both OSSEs based on high resolution simulations (50 m horizontal grid spacing) and a real data case study. In both experiments, radar data were either pre-gridded using two- or three-dimensional Cressman weighted average gridding techniques, or assimilated directly into the variational wind retrieval algorithm using the radar assimilation (RA) method. The outcomes of these experiments broadly confirmed the hypothesis that directly assimilating, rather than pre-gridding, radial velocity measurements results in more accurate 3D wind retrievals for the most challenging cases (\ie the supercell thunderstorms investigated here). The main findings from our study are summarized below:

\begin{itemize}
    \item Spatial interpolation has a very large effect on the quality of wind retrievals. Wind error magnitudes increased 300\% -- 350\% when using operational radar scanning geometries with realistic observational errors, compared to a control experiment with ``perfect'' observations defined everywhere within precipitating regions. 
    
    \item Spatial interpolation errors are greater for methods using pre-gridded radial velocities, resulting in average total error magnitudes of $\sim$7.3 m s$^{-1}$, compared with 6.2 m s$^{-1}$ in the RA experiments.

    \item If restricted to pre-gridded retrieval strategies, the 2D Cressman gridding performs similarly to the 3D Cressman method in terms of error scores (albeit less consistently), but more accurately resolves highly dynamic regions of the simulated storm due to the increased detail in radial velocity inputs.

    \item The RA method qualitatively resolves important dynamic features that the pre-gridded methods do not, including the mesocyclone, rear-flank downdraft, overturning circulation signatures and all major updraft trajectory pathways present within the modeled storm.
    
    \item The RA method also reduces errors in vertical vorticity, horizontal divergence and all three wind components (to an RMSE of $\sim$3.5 m s$^{-1}$ for $u$, $v$ and $w$), more accurately retrieves the maximum values of these quantities, and more reliably retrieves regions with intense updrafts/downdrafts as evidenced by greater Fractions Skill Scores.

    \item An investigation of a supercell case study showed promising agreement with the OSSE findings above, suggesting that the improvements observed for the RA method are applicable to real data in practice.
\end{itemize}


Future work will be aimed at generalizing these results using further OSSEs and real data examples containing different weather phenomena (\eg isolated convection, stratiform precipitation, and tropical cyclones). These studies should investigate whether the RA method shows comparable benefits to those described here for strong convection. We also aim to assess how the improvements achieved with the RA technique will function alongside other recent 3D wind retrieval developments, such as vertical vorticity constraints \citep{Shapiro09, Potvin12b} and spatially variable advection corrections \citep{Shapiro10a}. This is particularly relevant for the OSSE portion of our study, which assumed an instantaneous radar volume coverage pattern to control for the effects of non-stationarity. Finally, we aim to verify our findings against vertical velocity measurements from observed cases, as demonstrated recently in \citet{Gebauer22}.

\acknowledgments
The authors would like to acknowledge Robert Warren and three anonymous reviewers for revising and substantially improving this manuscript. JPB acknowledges industry research funding provided by Guy Carpenter and Company, LLC.

%
%
\datastatement
Radar data used in this study is available from the Australian Unified Radar Archive (\url{https://www.openradar.io/}). Model data is available upon request due to the considerable storage requirements required to house the high-resolution model outputs used in this study.

%

\appendix[A] 

\appendixtitle{Projection/Interpolation Operators}

The extensive use of pre-gridded radial velocities in recent wind retrieval publications \citep[\eg][]{North17, Dahl19, Oue19, Gebauer22}, and exclusive use in the open-source software packages MultiDop \citep{Lang17} and PyDDA \citep{Jackson19}, has obscured the once obvious distinction between pre-gridded and direct radar assimilation techniques \citep[\eg][]{Gao99, Rihan05}. In this section, we aim to clarify this distinction by providing a technical explanation of the projection and interpolation operators used in the RA method. 

\begin{figure}
\centering
\includegraphics[width=\columnwidth]{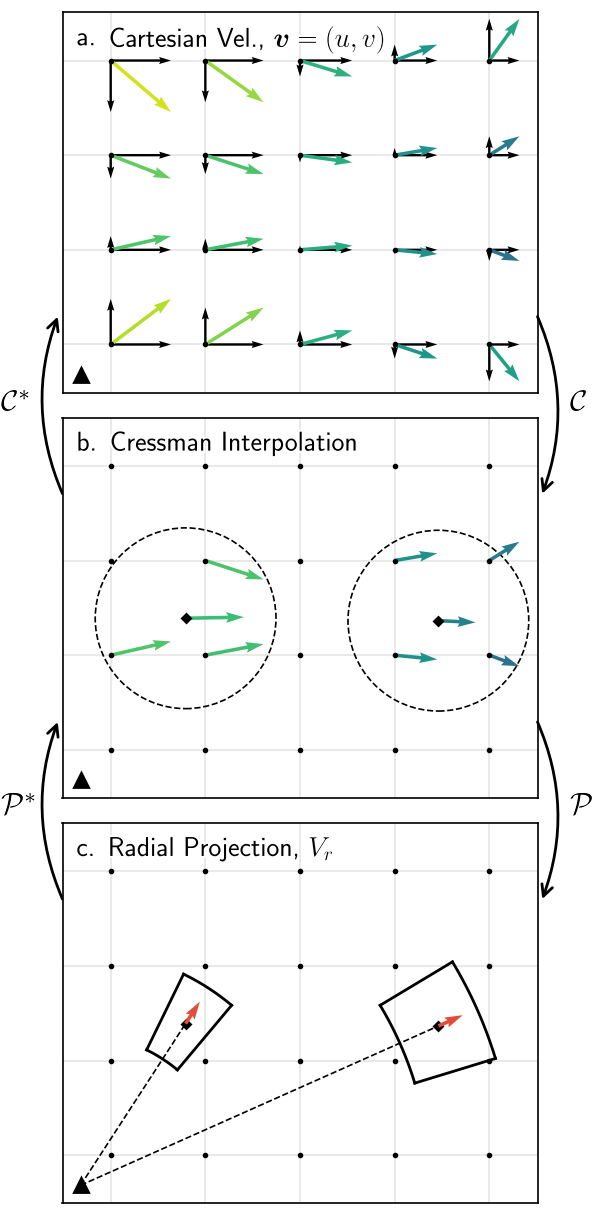}\\
 \caption{\label{interp_proj} An illustration of the Cressman interpolation ($\mathcal{C}$) and radial velocity projection ($\mathcal{P}$) operators and their adjoints ($\mathcal{C}^*$ and $\mathcal{P}^*$, respectively). (a) Colored arrows show a gridded, Cartesian velocity field (warmer colors represent larger magnitudes), with the individual velocity components shown in black. (b) Cressman interpolation from grid points within a radius of influence (dotted circles) of two radar observation points (diamonds). (c) Radial projection (red arrows) of the interpolated Cartesian velocities at the observation points. The black triangle indicates the radar position, and black circles and grid lines indicate the Cartesian analysis grid.}
\end{figure}

In the pre-gridded observational constraint in Eq. \ref{cart_obs}, the difference between the analysis fields and radar observations is a straightforward calculation on the analysis grid. However, when the observations are provided in their native observation geometry in the RA method, the comparison requires a forward operator to interpolate the analysis fields to the observation locations. Perhaps the simplest interpolation operator is a trilinear interpolation of the eight surrounding grid points \citep[\eg][B22]{Gao99,Gao04}. We tested this operator in our OSSE experiments (not shown) and found it performed poorly compared to the Cressman interpolation operator used throughout this study. We attribute these shortcomings to the propagation of observational noise and the introduction of first-order discontinuities into the analysis, which are characteristic of linear interpolation \citep[][B22]{Trapp00, Askelson00}. Wind retrievals are particularly sensitive to these effects due to their reliance on finite difference derivatives in the mass continuity equation \citep{Testud83}, meaning a Cressman interpolation operator (with the superior filtering qualities) is more suited to this application. The radius of influence in $\mathcal{C}$ was set to 1400 m for all of our experiments, having been experimentally optimized in a set of preliminary experiments (not shown), by setting it as a free variable in the Bayesian parameter optimization process described in Sec. \ref{results}\ref{retrieval_method}\ref{other_constraints}. As in \citet{Potvin12d}, we find our results are not particularly sensitive to the provision of this parameter.

We illustrate the implementation of the Cressman interpolation operator with a 2D Cartesian velocity field in Fig.\ \ref{interp_proj}a. Computationally speaking, the operator is set up prior to the analysis by noting which grid points are within a constant radius of influence from each observation (as in Fig.\ \ref{interp_proj}b), and the Cressman weights are pre-calculated according to Eq. \ref{cressman}. These weights are used to interpolate the analysis fields to the observation points, before the newly-interpolated Cartesian velocities are projected into radial velocities by the radial projection operator in Fig.\ \ref{interp_proj}c. Importantly, the $\mathcal{C}$ and $\mathcal{P}$ operators are not commutative, meaning the order in which they are applied is significant. The order presented here is more accurate as the radial projection takes place directly at the observation location, instead of attempting to interpolate a strictly non-smooth quantity in the reverse order. Also note that the Cressman interpolation in Fig.\ \ref{interp_proj}b is well defined, as more than one grid point will always be within a radius of influence from the observation points (due to the regularity of the Cartesian grid). Conversely, if one attempted to interpolate the radial velocity data in Fig \ref{interp_proj}c to the grid points with a similar radius of influence (as is attempted in pre-gridding methods), many grid points would have no valid observations, producing data voids in the result. The radius of influence must then be inflated to mask the data acquisition gaps (which are very large between constant elevation sweeps), thereby filtering important information from the resulting data field.

\appendix[B] 

\appendixtitle{Data Boundary Effects}

Dual-Doppler 3D wind retrievals are severely hampered by data coverage limitations that arise from two main sources: 1) weather radars only return reliable information within regions with backscatterers (usually within clouds), and 2) radars only take measurements at set locations defined by their scanning strategy, meaning there are significant data gaps even within strong echo regions. We showed in our control OSSE experiments that eliminating the latter error source (by assigning valid data everywhere within echo regions) reduced total errors by more than  300\% compared to a standard operational scanning pattern. Furthermore, if error source (1) is eliminated (by assigning valid radial velocity data everywhere, not shown), the retrieval errors drop to near-zero. However, when both error sources are present (as is the case in real data), care must be taken to ensure boundary errors do not propagate into valid data regions. Here, we detail how imposing boundary conditions on the observational constraint can mitigate these effects.

\begin{figure*}
\centering
\includegraphics[width=\textwidth]{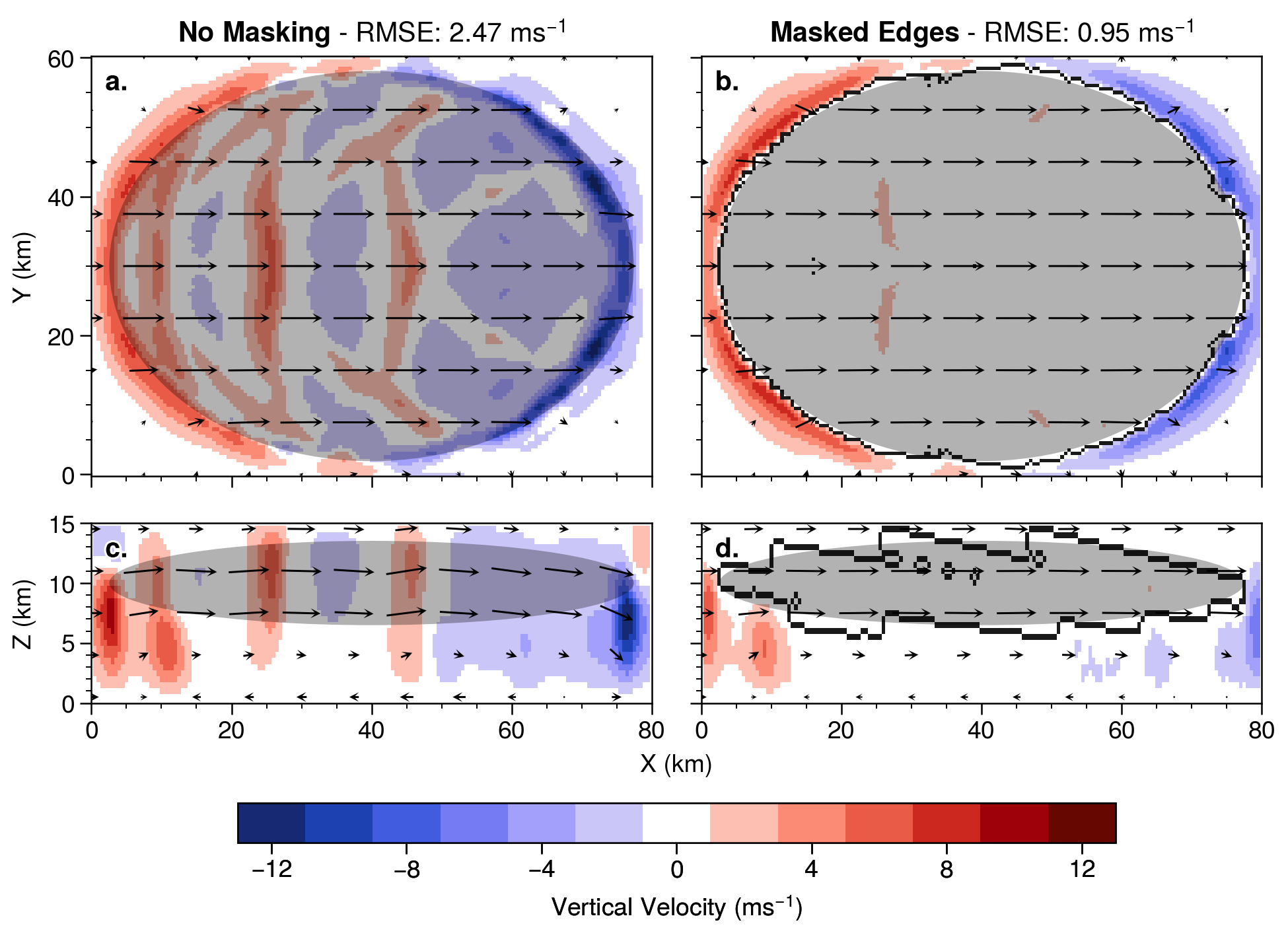}\\
 \caption{\label{boundary} Wind retrievals from the simple observing system simulation experiment described in Appendix B. Horizontal cross sections at Z = 10 km are shown in (a) and (b), along with vertical cross sections at Y = 30 km in (c) and (d). The radar echo region is highlighted in grey, while vertical winds are shaded and arrows show wind velocities along the cross section plane. Pixels containing radar observation boundaries are colored black in (b) and (d).}
\end{figure*}

Firstly, the errors introduced at data boundaries may be best understood by considering an isolated, high-altitude data point with strong horizontal velocity, and no vertical velocity. If the radar is situated parallel to the strong horizontal velocities, the measured radial velocity will be large -- such that when it is projected back into Cartesian coordinates, a significant portion will be interpreted as vertical velocity (especially for higher elevation angles). In the absence of valid radial velocities in the surrounding grid points, the mass continuity and smoothing constraints will then construct a spurious updraft/downdraft region to match the isolated observation point, which is what we observed in isolated regions in Fig.\ \ref{horiz_plot}a, and along data boundaries in other studies \citep[\eg][]{Collis10}. We have found that setting the vertical gradient of the observational constraint equal to zero at these boundary grid points (boundaries are defined as neighbors to a data void), significantly reduces these errors, and we have designed a simple OSSE to demonstrate these effects.

The simple OSSE contains purely westerly flow ($V = W = 0$), which we have derived to mimic the thought experiment above. To this end, we fit a fourth order polynomial to the average $u$ profile from the ARPS simulation (containing an upper-level jet at $\sim$10 km): $U = 0.0021z^4 - 0.067z^3 + 0.30z^2 + 5.14z + 0.46$, where $z$ is in km. The domain is the same as in our model OSSEs ($80 \times 60 \times 15$ km in the $x,y,z$ dimensions, with an isotropic 500 m grid spacing), and the radars are positioned roughly parallel to the westerly flow at position 7 in Fig.\ \ref{setup}. We limit the radar echo regions to an elliptical shape spanning the entire domain horizontally, with a depth of roughly 6 km, centered at a height of 10 km. We then simulated radial velocities as in Section 3a, and finally retrieved vertical winds using the radar assimilation retrieval method both without and with boundary masking (Figs. \ref{boundary}a, \ref{boundary}c and Figs. \ref{boundary}b, \ref{boundary}d, respectively).

While the westerly horizontal flow is retrieved well in the experiment with no boundary masking, significant spurious vertical motions are created within the valid data region (illustrated by the gray shading). These spurious features within the cloud are associated with the aforementioned data boundary effects, while the strongest updrafts/downdrafts at the horizontal edges of the cloud are induced by the mass continuity constraint at the domain edges (which also creates spurious return flow near the surface in Fig.\ \ref{boundary}c). In Figs. \ref{boundary}b and \ref{boundary}d, the $w$ component of the observational adjoint is set to zero (\ie $\mathcal{C}^*\mathcal{P}^*V_r = 0$) at the boundary points (plotted in black). Note that the spurious updraft features within the cloud are almost completely eliminated, leaving only those outside the valid data region associated with mass continuity. The total RMSE (within the valid data region) drops from 2.47 m s$^{-1}$ to 0.95 m s$^{-1}$ as a result of this computational procedure. We have noted a similar elimination of spurious updrafts in the model OSSEs throughout this study.

\bibliographystyle{ametsocV6}
\bibliography{references_winds_ams.bib}

\end{document}